\documentclass[aps,prd,amsmath,twocolumn,showpacs]{revtex4}

\usepackage{epsfig}
\usepackage{graphics}
\usepackage{latexsym}
\usepackage{amsmath}
\usepackage{amssymb}
\usepackage{rotating}
\usepackage{subfigure}
\usepackage{bm}
\usepackage{color}
\usepackage{ulem}

\usepackage[colorlinks = true,
linkcolor = magenta,
urlcolor  = blue,
citecolor = red,
anchorcolor = blue]{hyperref}

\begin{document}

\title{ First determination of $D^{*}$-meson fragmentation functions and their uncertainties at next-to-next-to-leading order }

\author{Maryam Soleymaninia$^{1,4}$}
\email{maryam\_soleymaninia@ipm.ir}

\author{Hamzeh Khanpour$^{2,4}$}
\email{Hamzeh.Khanpour@mail.ipm.ir}

\author{S. Mohammad Moosavi Nejad$^{3,4}$}
\email{mmoosavi@yazd.ac.ir }

\affiliation {
$^{1}$Department of Physics, Shahid Rajaee Teacher Training University, Lavizan, Tehran 16788, Iran          \\
$^{2}$Department of Physics, University of Science and Technology of Mazandaran, P.O.Box 48518-78195, Behshahr, Iran    \\
$^{3}$Faculty of Physics, Yazd University, P.O.Box 89195-741, Yazd, Iran                                    \\
$^{4}$School of Particles and Accelerators, Institute for Research in Fundamental Sciences (IPM), P.O.Box 19395-5531, Tehran, Iran       
 }

\date{\today}

%
\begin{abstract}

We present, for the first time, a set of next-to-next-to-leading order (NNLO) fragmentation functions (FFs) describing the production of charmed-meson $D^{*}$ from partons. Exploiting the universality and scaling violations of FFs, we extract the NLO and NNLO FFs through a global fit to all relevant data sets from single-inclusive $e^+e^-$ annihilation. The uncertainties for the resulting FFs as well as the corresponding observables are estimated using the Hessian approach.
We evaluate the quality of the {\tt SKM18} FFs determined in this analysis by comparing with the recent results in literature and show how they describe the available data for single-inclusive $D^{*\pm}$-meson production in electron-positron annihilation.
As a practical application, we apply the extracted FFs to make our theoretical predictions for the scaled-energy distributions of $D^{*\pm}$-mesons 
inclusively produced in top quark decays. We explore the implications of {\tt SKM18} for LHC phenomenology and show that our findings of this study can be introduced as a channel to indirect search for top-quark properties.

\end{abstract}

\maketitle
\tableofcontents{}

%
\section{Introduction}\label{sec:introduction}
%

Fragmentation functions (FFs)~\cite{Bertone:2017tyb,Ethier:2017zbq,Anderle:2015lqa,MoosaviNejad:2016qdx} describe the nonperturbative part of hard-scattering processes and along with the
parton distribution functions (PDFs) of initial hadrons (in hadron-hadron collision) and
parton-level differential cross sections are three necessary ingredients to obtain theoretical predictions for hadroproduction cross sections~\cite{Nocera:2017zge,Hou:2017khm,Eskola:2016oht,Frankfurt:2016qca,Goharipour:2017uic}.

Studies over the past two decades have provided valuable important information on the structure of hadrons. The FFs and PDFs extracted from these studies encode the long-distance dynamics of the interactions among quarks and gluons so the partonic cross sections encode the short-distance dynamics of the interactions. The process-independent FFs, $D_i^H(z, \mu_F^2)$, describe the probability for a parton $\it{i}$ at the factorization scale $\mu_F$ to fragment into a hadron $\it{H}$ carrying away a fraction $\it{z}$ of its momentum.
The specific importance of FFs is for their model-independent predictions of the cross sections (or decay rates) at colliders where the observables involving identified hadrons are detected in the final state~\cite{Slovak:2017xdc,Anderle:2017cgl,Ethier:2017zbq}. The knowledge of FFs has received quite some interest in, for example, tests of quantum chromodynamics (QCD) such as theoretical calculations for recent measurements of inclusive production in proton-proton collisions at the Relativistic Heavy Ion Collider (RHIC) and the CERN LHC, and in investigating the origin of the proton spin to describe the internal structure of nucleons.

In the naive parton model, the nonperturbative FFs are independent of the factorization scale $\mu_F$, but in the QCD-improved parton model 
their scaling violations are subject to the perturbatively computable Dokshitzer-Gribov-Lipatov-Alteralli-Parisi (DGLAP) evolution equations~\cite{DGLAP}.
In this respect, the PDFs and FFs are on the same footing. Like PDFs, the FFs must be extracted from experimental data through global QCD analyses, possibly from a variety of hard-scattering processes~\cite{Bertone:2017tyb,Ethier:2017zbq,Anderle:2015lqa,Zarei:2015jvh,Boroun:2015aya,Boroun:2016zql}.
The FFs are included in hadrproduction processes such as electron-positron Single-Inclusive Annihilation (SIA), lepton-hadron Semi-Inclusive Deep-Inelastic Scattering (SIDIS) and hadron-hadron scattering processes. Information obtained from SIDIS multiplicities and from hadron-hadron collisions is particularly useful in order to achieve a complete flavor decomposition into quark and antiquark FFs along with a direct determination of the gluon FF.
Among them, SIA remains the theoretically cleanest process to extract the fragmentation densities since its interpretation does not require the simultaneous knowledge of PDFs~\cite{Anderle:2017cgl}. Recent progresses in the extraction of FFs have been focused on light charged mesons (pions and kaons), for which data are more numerous as they dominate the identified hadron yields. For example, in Ref.~\cite{Soleymaninia:2013cxa} we have determined the $\pi^\pm$ and $K^\pm$ FFs,
both at leading-order (LO) and next-to-leading order (NLO) accuracy in perturbative QCD through a global QCD fit to the SIA data and the SIDIS asymmetry data from
HERMES and COMPASS. 
There, we have broken the symmetry assumption between the quark and antiquark FFs for favored partons by using the asymmetry data. In~\cite{Nejad:2015fdh}, we have also presented sets of proton FFs with their corresponding uncertainties through a global fit to all relevant data sets of SIA, taking the finite-mass effects of proton into account. Other well-known groups which have determined sets of NLO FFs for these two mesonic species are: AKK~\cite{Albino:2008fy}, LSS~\cite{Leader:2015hna}, DSEHS~\cite{deFlorian:2014xna},  
DSS~\cite{deFlorian:2007aj,deFlorian:2007hc}, HKNS~\cite{Hirai:2007cx} and JAM~\cite{Sato:2016wqj} collaborations who have used different phenomenological models and  variety of experimental data. In these studies, main focus was put on quantifying the effects of the inclusion of new measurements on the FFs, although in the JAM and HKNS fits these were limited to SIA data.
Apart form the different data set used in their determinations, these collaborations have also introduced some methodological and theoretical improvements over previous determinations, specifically, in order to achieve a more reliable estimate of the uncertainties of FFs. For example, the iterative Hessian procedure~\cite{Pumplin:2000vx} has been used in the DSEHS analysis, while the iterative Monte Carlo approach~\cite{Sato:2016tuz} has been developed in the JAM analysis. 
Recently, in  Refs.~\cite{Bertone:2017tyb,Ball:2017nwa} using  data from  a comprehensive set of SIA measurements, authors have presented a new determination of the FFs of charged pions, charged kaons, and protons/antiprotons at next-to-next-to-leading  order (NNLO) in perturbative QCD. They resulted that the systematic inclusion of higher-order QCD corrections significantly improves the description of the data, especially in the small-$z$ region. 

Generally, the theoretical treatment of heavy quarks provides a unique laboratory to test perturbative QCD. In fact, heavy flavor cross sections which have been measured both at very high energies at the LHC and at various low energy experiments poses unique challenges to our deep understanding of QCD. Specifically, charm production
cross sections are applied, for instance, to further constrain the gluon PDF at small-$x$~\cite{Gauld:2015yia}, and they play a vital role in
neutrino astrophysics and cosmic-ray~\cite{Garzelli:2015psa,Gauld:2015kvh}. Another importance of charm production cross sections concerns the modification of heavy
flavor yields in heavy-ion collisions where highly energetic partons can traverse the quark-gluon plasma thereby attaining valuable information about the properties
of the QCD medium. For more detail, see Ref.~\cite{Kang:2016ofv}.

In~\cite{Kneesch:2007ey}, using the SIA data from the Belle, CLEO, ALEPH and OPAL Collaborations, authors have determined nonperturbative charmed-meson FFs at NLO in the modified Minimal-Subtraction ($\overline{MS}$) factorization scheme.  To extract the results, the General-Mass Variable-Flavor-Number scheme (GM-VFNs) was adopted. However, the most recent Belle data which were included in their analysis have been removed due to an unrecoverable error in the measurement. Recently, in~\cite{Anderle:2017cgl} authors have calculated the FFs of charged $D^*$-meson at next-to-leading order accuracy using the available data for single-inclusive $D^*$-meson
production in $e^+e^-$ annihilation, hadron-hadron collisions, and in proton-proton scattering. While quark-to-hadron FFs can be relatively well constrained from SIA data alone, the gluon FF can be well constrained via proton-proton scattering data. It should be noted that, to compute the NLO FFs authors have used the Zero-Mass Variable-Flavor-Number scheme (ZM-VFNs)~\cite{Binnewies:1998vm} where the heavy quark masses are set to zero in all partonic cross sections so the heavy quarks are treated as the
active, massless partons in the proton and the nonzero values of the charm- and bottom-quark masses only enter through the initial conditions of the FFs. This scheme is applicable for sufficiently high energies and transverse momenta.

In the present work, using the ZM-VFN scheme we focus on the hadronization of charm- and bottom-quarks into $D^{*\pm}$-mesons, which are of particular relevance in the era of the LHC.
We will provide the first global QCD analysis of $(c,b)\to D^{*\pm}$ FFs at NNLO accuracy through a global  QCD fit to SIA data from ALEPH~\cite{Barate:1999bg} and OPAL~\cite{Ackerstaff:1997ki} collaborations at LEP. Our analysis which is named as  ``{\tt SKM18}'' from now on, is limited to SIA data only due to the lack of other single-inclusive particle
production cross sections at NLO and NNLO accuracy. We will also present an attempt to estimate the uncertainties of the extracted FFs as well as the resulting normalized total cross sections, for which we adopt the Hessian method~\cite{Martin:2002aw,Pumplin:2001ct,Martin:2009iq}.

As it is well-known, in the Standard Model of particle physics the top quark has a short lifetime so it decays before hadronization takes place. In fact, at the lowest order the top quark  decays as $t\to W^+b$ followed by $b\to X+Jets$, where $X$ refers to the detected hadrons in the final state.  Therefore, at the LHC the study of energy spectrum of produced hadrons through top decays is proposed as a new channel to indirect search for the top quark properties. As an example of a possible application, the extracted FFs in our analysis are used to make the theoretical predictions for the energy distributions of $D^{*\pm}$-mesons in polarized and unpolarized top decays. We shall compare our results with the ones obtained through the NLO FFs extracted in~\cite{Kneesch:2007ey}.

This paper is organized as follows. In Sec.~\ref{sec:QCD analysis} we explain the QCD analysis of hadronization process in electron-positron annihilation by introducing FFs. We describe our formalism and parametrization form at the initial scale for $D^{*\pm}$ FFs in Sec.~\ref{sec:parametrization}.
In Sec.~\ref{sec:data selection}, we discuss  the related experimental data and their effects in our global analysis. In following, the minimization method and the Hessian uncertainty approach to calculate the errors in {\tt SKM18} analysis are present in Sec.~\ref{sec:errorcalculation}.
In Sec.~\ref{sec:Results}, the full results for the $D^{*\pm}$-FFs and their uncertainties are listed. We also present a comparison of {\tt SKM18} results with experimental data and the other models in this section. 
The theoretical uncertainties due to the variation of the renormalization and factorization scales at the NLO and NNLO accuracies are studied at the end of this section. 
Our predictions for the energy spectrum of charmed mesons produced in top quark decays are presented in Sec.~\ref{sec:top decay}. Our result and conclusion are summarized in Sec.~\ref{sec:conclusion}.

%
\section{QCD Analysis Framework up to NNLO accuracy}\label{sec:QCD analysis}
%

The optimal way to determine the $D^{*}$-FFs is to fit them to experimental date extracted from the single-inclusive $e^{+}e^{-}$ annihilation processes
\begin{eqnarray}\label{process}
e^{+}e^{-}\rightarrow (\gamma ,Z)\rightarrow D^{*}+X,
\end{eqnarray}
where $X$ stands for the unobserved final jets. Since, this has less contributions from background processes in comparison with the hadron collisions, then one does not need to deal  with the uncertainties introduced by PDFs. 
In the following, we explain how to evaluate the cross section of the above process in the parton model of QCD within the ZM-VFN scheme. Note that, in the ZM-VFNS (or massless scheme), the number of active quark flavors also increases with the flavor thresholds.

If we denote the four-momenta of the exchanged virtual gauge boson and the $D^{*}$ meson by $q$ and $p_D$, so that $s=q^2$ and $p_D^2=m_D^2$, the scaling variable $x_D$ is defined as $x_D=2(p_D\cdot q)/q^2$, similarly to the Bjorken-$x$ variable in deep inelastic scattering. In the center-of-mass frame, this variable is reduced to $x_D=2E_D/\sqrt{s}$ which refers to the energy of $D^{*}$-meson in units of the beam energy.

According to the factorization theorem of the improved QCD-parton model~\cite{Collins:1998rz}, the differential cross section of  process   \eqref{process} can be written as a convolutions of perturbatively calculable partonic cross sections $d\sigma_i(x_i, \mu_R, \mu_F)/dx_i$ with the $D^{D^{*}}_i (z, \mu_F^2)$-FFs, where $i$ stands for one of the fragmenting partons $i=g, u,\bar u, \cdots, b,\bar b$. This convolution reads
\begin{eqnarray}\label{convolution}
\frac{1}{\sigma_{\rm tot}}\frac{d}{dx_D} \sigma(e^+e^-\rightarrow D^*X)&=& \nonumber \\
&&\hspace{-4.5cm} \sum_i \int_{x_D}^1\frac{dx_i}{x_i} D_i^{D^{*}}(\frac{x_D}{x_i}, \mu_F)\frac{1}{\sigma_{\rm tot}}\frac{d\sigma_i}{dx_i}(x_i, \mu_R, \mu_F).
\end{eqnarray}
The renormalization ($\mu_R$) and factorization ($\mu_F$) scales are arbitrary quantities, however, a choice often made consists of setting $\mu^2 = \mu_F^2 = Q^2$
and following Ref.~\cite{Anderle:2017cgl} we shall adopt this convention in  this work. The variable $x_i$ is also defined in analogy to $x_D$ as $x_i=2(p_i\cdot q)/q^2$, where $p_i$ is the four-momentum of parton $i$.
To compare our theoretical calculations with experimental data, we normalized the differential cross section $d\sigma/dx_D$ to the total cross section up to NNLO for $e^+e^-$ annihilation into hadrons ($\sigma_{\rm tot}$) which reads
\begin{eqnarray}\label{sigmatotal}
\sigma_{\rm tot}&=&\frac{4\pi\alpha^2(Q)}{Q^2}\bigg(\sum_i^{n_f}\tilde{e}_i^2(Q)\bigg) (1+\alpha_s K_{\rm QCD}^{(1)} \nonumber \\
&&+\alpha_s^2K_{\rm QCD}^{(2)} + \cdots),
\end{eqnarray}
where $\alpha$ and $\alpha_s$ are the electromagnetic and the strong
coupling constants, respectively, and the coefficients $K_{\rm QCD}^{(i)}$ (with $K_{\rm QCD}^{(1)}=3C_F/4\pi$) indicate the  perturbative QCD corrections to the LO result and are currently known up to ${\cal O}(\alpha^3_s)$~\cite{Gorishnii:1990vf}. It should be noted that, to calculate the NLO or NNLO FFs one needs to have the partonic cross sections $d\sigma_i/dx_i$ (Wilson coefficients) at each desired order. The NNLO Wilson coefficients are analytically presented in Refs.~\cite{Rijken:1996vr,Rijken:1996ns,Mitov:2006wy}.

%
\section{Phenomenological parametrization up to NNLO } \label{sec:parametrization}
%

In order to choose the best parametrization for {\tt SKM18} global analysis of $D^{*\pm}$-FFs,   we tested different models
such as Peterson~\cite{Kniehl:2006mw, Binnewies:1997xq} and Bowler~\cite{Kneesch:2007ey} and other simple forms applied for light hadron FFs (i.e. pion, kaon and proton)~\cite{Hirai:2007cx}. 
Finally, we adopted the Bowler parametrization form because of its best fit to low number of data for $D^{*\pm}$-meson. 
Therefore, we parametrize the $z$ distributions of the $c(\bar{c})$ and $b(\bar{b})$  quark FFs at their starting scales $\mu_0$ as suggested by Bowler, i.e.  
\begin{eqnarray}
\label{parametrization}
D^{D^{*\pm}}_i(z, \mu _0^2) = N_iz^{-(1 + \alpha _i ^2)}(1 - z)^{\beta _i} e^{-\alpha _i ^2/z},
\end{eqnarray}
with three free parameters; N, $\alpha$ and $\beta$. Our fitting procedure is going as follows. At the scale $\mu_0$, the charm and bottom quark FFs are taken to be of the Bowler form as in Eq.~\eqref{parametrization}, while the FFs of gluon and light quarks ($q=u, d, s$) are set to zero, i.e. 
\begin{equation}
D^{D^{*\pm}}_i(z,\mu _0^2)=0,~~~~ i=u,\bar{u},d,\bar{d},s,\bar{s},g.
\end{equation}
Then, these light and gluon FFs are evolved to higher scales using the DGLAP evolution equations~\cite{DGLAP} at NLO or NNLO.  To proceed, we set the initial parametrization scale as $\mu _0^2 =18.5$~GeV$^2$ which is a little grater than the b-quark threshold, i.e. $Q^2=m_b^2=(4.3)^2$~GeV$^2$.  The advantage of taking this value for the initial scale is due to the fact that the time-like matching conditions are currently known only up to NLO accuracy and with this input scale the heavy-quark thresholds should not be crossed in the QCD evolution.  

Since all hadrons in electron-positron annihilation originate from the produced $q \bar{q}$ pair, the multiplicities for $D^{*+}$ and $D^{*-}$ are the same and the experimental data for charged hadrons are usually presented for their sum, i.e.
\begin{eqnarray}
\label{+-}
\frac{1}{\sigma_{tot}}\frac{d\sigma ^{D^{*\pm}}}{dz}=\frac{1}{\sigma_{tot}}\frac{d\sigma ^{D^{*+}}}{dz}+\frac{1}{\sigma_{tot}}\frac{d\sigma ^{D^{*-}}}{dz} \,.
\end{eqnarray}
According to the parton structure of $D^{*-}$ ,  the FFs of $D^{*-}$ can be obtained as
\begin{equation}
D^{D^{*-}}_q(z,\mu ^2)=D^{D^{*+}}_{\bar{q}}(z,\mu ^2) \,,
\end{equation}
and for the gluon FF, it reads
\begin{equation}
D^{D^{*-}}_g(z,\mu ^2)=D^{D^{*+}}_g(z,\mu ^2) \,.
\end{equation}

It seems that one should determine six free parameters for the FFs of charm and bottom quarks into the $D^{*\pm}$-meson using experimental data. Since, $N_c$ can not be constrained by data, then one should fix it from the beginning so the actual parameters which should be determined are five parameters.
In the next section, we will discuss how further data collections are required to determine all six parameters described above.

%
\section{Experimental input} \label{sec:data selection}
%

Most of experimental data for $D^{*\pm}$ in $e^+e^-$ annihilation is reported by ALEPH~\cite{Barate:1999bg}, OPAL~\cite{Ackerstaff:1997ki}, CLEO~\cite{Artuso:2004pj} and Belle~\cite{Seuster:2005tr} Collaborations.
An overview of the data included in {\tt SKM18} global analysis of FFs is presented in Tables.~\ref{Data-Set_NLO} and ~\ref{Data-Set_NNLO} for the total, $c$-tagged and $b$-tagged SIA cross sections from ALEPH~\cite{Barate:1999bg} and OPAL~\cite{Ackerstaff:1997ki} Collaborations. For each dataset we include the corresponding published reference, the number of data points in the {\tt SKM18} NLO and NNLO $D^{*\pm}$-FFs determinations, and also the center-of-mass energy. The total number of data points for the {\tt SKM18} FFs determination is 59 at NLO and NNLO.
ALEPH and OPAL Collaborations at LEP  present their experimental data at $Q = M_Z$, the mass of the $Z$ boson, while Belle and CLEO provide their data in lower energy, i.e. $Q = 10.5$ GeV.
In this range of energy, all $D^{*\pm}$ in $e^+e^-$ annihilation coming from bottom decays are excluded because they are bellow the b-quark mass threshold.

In spite that the Belle data has been published in Ref.~\cite{Belle}, but results are not yet available  on the HEPDATA web page. While KKKS08 group~\cite{Kneesch:2007ey} have used the Belle data in their analysis of FFs, on the HEPDATA web page it is explained that ``due to an unrecoverable error in the measurement, the data for this record have been removed by the request of the authors"~\cite{Belle}. Therefore, we are not able to include this data set while KKKS08 group applied them. This makes the  KKKS08 analysis unreliable.

Note that, CLEO present the data as distributions $d\sigma/dx_p$ for hadron $H$ where the scaled momentum is defined as
\begin{equation}\label{xp}
x_p= p/p_{max}=\sqrt{(x^2-\rho _H)/(1-\rho _H)} \,,
\end{equation}
where $\rho _H=4m^2_H/s$ in which $m_H$ stands for the hadron mass and the allowed values of $x_p$ are $0\leq x_p\leq 1$. Also the conversion formula for differential cross section reads
\begin{equation}\label{xp+cross}
\frac{d\sigma}{dx_p}(x_p)=(1-\rho _H)\frac{x_p}{x}\frac{d\sigma}{dx}(x),
\end{equation}
where $x=\sqrt{(1-\rho_H)x_p^2+\rho_H}$ \cite{Kneesch:2007ey}.
	
Ignoring the hadron mass leads to $(d\sigma /dx_p)(x_p)=(d\sigma/dx)(x)$. This assumption  leads to a large difference between theory  and experimental
results. 
When we apply the CLEO data in our analysis, it disorganizes our results and also the $\chi ^2$ increases.
While this data set has been included in the extractions of $D^*$-FFs  in~\cite{Kneesch:2007ey,Cacciari:2005uk}, it has not been included in recent analysis by AKSRV17 group~\cite{Anderle:2017cgl} and authors found noticeable tensions between the CLEO and ALEPH data. In conclusion, we do not include the CLEO low-energy SIA data in {\tt SMK18} analysis.

The ALEPH and OPAL experimental data sets~\cite{Barate:1999bg,Ackerstaff:1997ki} present the total, charm and bottom flavor tagged for the cross section distributions normalized to the total hadronic cross section and include the branching fractions  of the decays  used to identify the $D^{*\pm}$ mesons, namely $D^{*+}\rightarrow D^0 \pi ^+ $ followed  by $D^0 \rightarrow K^-\pi ^+$.
Therefore, we divide these two data sets by the respective decay branching fractions given by~\cite{Patrignani:2016xqp}, i.e. $B(D^{*+}\rightarrow D^0 \pi ^+ )=(67.7\pm 0.5)\%$ and $B(D^0 \rightarrow K^-\pi ^+)=(3.93\pm0.04)\%$. Unfortunately, the $c$-tagged experimental data from ALEPH Collaboration are presented in graphical form, and we can not reach the numerical values~\cite{Barate:1999bg}.  The normalized $c$-tagged cross sections from OPAL experiment~\cite{Ackerstaff:1997ki} are included in the {\tt SKM18} NLO and NNLO QCD fits.

AKSRV17 group~\cite{Anderle:2017cgl} include electron-positron annihilation, hadron-hadron collisions and hadron-in-jet data in their global analysis of $D^*$-FFs at NLO. Since, using different hadronization processes increase the number of data, the $c$-tagged cross section can be further constrained. Also, the AKSRV17 analysis allows for a nonzero gluon FF at the initial scale in order to get a good global fit of all different hadronization processes data.
Note that the Wilson coefficient functions are only known at NNLO just for SIA process~\cite{Rijken:1996vr,Rijken:1996ns,Mitov:2006wy}. Since our global analysis is performed at NNLO, which is the main purpose of  this study, we limit the potential of global determination of FFs to $e^+e^-$ annihilation.

In our analysis, we apply the massless scheme where we ignore the finite-mass effects of heavy quarks as well as the charmed-meson mass effects. The cuts applied to SIA data are as follows. We exclude some of the SIA data in our global analysis. The LEP data (ALEPH and OPAL) taken at $Q = M_Z$ are used only in the interval $0.1 < z < 0.95$.

%
\section{The calculation method of errors } \label{sec:errorcalculation}
%

In {\tt SKM18} global analysis, the total $\chi^2$ is calculated in comparison with the experimental data for $D^{*\pm}$ production in $e^+e^-$ annihilation.
For calculating the total $\chi^2$, the theoretical functions should be obtained at the same experimental $z$ and $\mu^2$ points. 
As we explained, the $\mu^2$ evolution is calculated by the well-known DGLAP evolution equations~\cite{DGLAP} and the total cross section for $e^+e^-$ annihilation into hadrons ($\sigma_{\text {tot}}$) is obtained from Eq.~\eqref{sigmatotal}. The simplest and fastest method to calculate the total $\chi^2(\{\eta_i\})$ for independent sets of unknown fit parameters $\{\eta_i\}$ reads

\begin{eqnarray}\label{eq:chi2-simple}
\chi^{2} (\{\eta_i\}) =
\sum_{i}^{n^{\text{data}}} \frac{({\cal O}^{\rm data}_i
- T^{\text{theory}}_i (\{\eta_i\}))^2} {(\sigma^{\text{data}}_i)^{2}} \,,
\end{eqnarray}

where ${\cal O}^{\text{data}}_{i}$ stand for the experimental values related to the desired observables and $T^{\text{theory}}_{i}$ are the corresponding theoretical values of  $D^{*\pm}$ productions at the same experimental $z$ and $\mu^2$ points. The experimental errors $\sigma^{\text{data}}_{i}$ are calculated from statistical and systematic errors added in quadrature, $(\sigma^{\text{data}})^{2}=(\sigma^{\text{sys}})^{2} + (\sigma^{\text{stat}})^{2}$. 
The unknown parameters of FFs, $D^{D^{*\pm}}_{i}(z, \mu_0^2)$, presented in Eq.~\eqref{parametrization} are determined so as to obtain the minimum $\chi^{2}$. The optimization of a given function is done by the CERN program {\tt MINUIT}~\cite{James:1994vla}.

In order to illustrate the effects arising from the use of $D^{*\pm}$ production data sets in our analyses, in Tables.~\ref{Data-Set_NLO} and \ref{Data-Set_NNLO}, we show the $\chi^{2}$ for each data sets. These tables clearly illustrate the quality of our QCD fits at NLO and NNLO accuracy in terms
of the individual $\chi^{2}$-values obtained for each experiment. The total $\chi^{2}/{\text{d.o.f}}$ for the resulting fit to the $D^{*\pm}$ productions are $1.31$ and $1.27$ for our NLO and NNLO fits, respectively.

Since most of experiments usually come with an additional information on the fully correlated normalization uncertainty, $\Delta{\cal N}_{n}$, the simple $\chi^{2}$ definition presented in Eq.~\eqref{eq:chi2-simple} needs to be corrected in order to account for the corresponding normalization uncertainties.
In that case and in order to determine the best fit parameters, we need to minimize the $\chi^{2}_{\text{global}}(\{\eta_i\})$ function in respect to the free unknown parameters.
Considering the discussion mentioned above, the $\chi^{2}_{\text{global}}(\{\eta_i\})$ function can be written as,

\begin{equation}\label{eq:chi2}
\chi_{\text{global}}^{2} (\{\eta_{i}\}) =
\sum_{n=1}^{n^{\text{exp}}} \, {\cal W}_{n} \chi_{n}^{2}\,,
\end{equation}

where ${\cal W}_{n}$ is a weight factor for the $n^{\text{th}}$ experiment and

\begin{eqnarray}\label{eq:chi2globalanalysisFFs}
\chi_{n}^{2} (\{\eta_i\})&=&
\left(\frac{1-{\cal N}_{n}} {\Delta{\cal N}_{n}}\right)^{2} + \nonumber \\
&&\hspace{-0.5cm}\sum_{k=1}^{N_n^{\text{data}}}
\left(\frac{({\cal N}_{n}  \,
{\cal O}_k^{\text{data}}-{T}_k^{\text{theory}}
(\{\eta_i\})}{{\cal N}_{n} \,
\delta{D}_{k}^{\text{data}}}\right)^{2}\,,
\end{eqnarray}

in which $n^{\text{exp}}$ refers to a given individual experimental data set and $N^{\text{data}}_{n}$ corresponds to the number of data points in each data set.

The normalization factors $\Delta{\cal N}_{n}$ in Eq.~\eqref{eq:chi2globalanalysisFFs} can be fitted along with the fitted parameters $(\{\eta_i\})$ of Eq.~\eqref{parametrization} and then can keep fixed. The obtained normalization factors $\Delta{\cal N}_{n}$ are also presented in Tables.~\ref{Data-Set_NLO} and \ref{Data-Set_NNLO} for our NLO and NNLO analyses.

The determination of nonperturbative FFs through QCD fits to the data is a statistical procedure that necessarily implies a variety of assumptions.
The most important one is the input parameterization functions of the charmed-meson FFs and the propagation
of the experimental uncertainties into them~\cite{Metz:2016swz,Bertone:2017tyb,Ethier:2017zbq,Anderle:2016czy,Anderle:2015lqa,deFlorian:2017lwf}.
The assessment of uncertainties of PDFs and the corresponding observables has seen significant progress in recent QCD analyses~\cite{,Hou:2017khm,Eskola:2016oht,Gao:2017yyd,Ball:2017nwa}.
Among the different approaches in literature, the ``Hessian method''~\cite{Martin:2002aw,Pumplin:2001ct,Shoeibi:2017lrl}, the Lagrange multiplier (LM) technique~\cite{Martin:2009iq} and newly Neural Network (NN) in which the NNPDF uses~\cite{Ball:2016neh,Ubiali:2014bva,Ball:2013tyh,Nocera:2014gqa} are the most reliable ones.

The well-known and practical method is ``Hessian method'', which has been widely used to extract the uncertainties of the PDFs, polarized PDFs and nuclear PDFs as well as the corresponding observables in our previous analyses.~\cite{Arbabifar:2013tma,Khanpour:2017fey,Khanpour:2017cha,MoosaviNejad:2016ebo,Khanpour:2016uxh,Khanpour:2016pph,Shoeibi:2017zha,Shahri:2016uzl}. Although, the technical details of the Hessian method are described in the mentioned references, a short summary of the method is explained here.

As we already discussed, our method is the $\chi_{n}^{2}(\{\eta_{i}\})$ fitting procedure in the global QCD analyses. For the determination of FFs uncertainties, we have used the well-known ``Hessian'' or error matrix approach.
This approach confirms that the {\tt SKM18} fitting methodology used in this QCD analysis can faithfully reproduce the input charmed-meson FFs in the region where the $D^{*+}$ productions data sets are sufficiently constraining.
The fit parameters are denoted as $\eta_{i}$ ($i$ = 1, 2, ..., $N$), where $N$ refers to the total number of the fitted parameters. One can expand the $\chi^{2}$ around the minimum $\chi^{2}$ point, i.e. $\hat \eta$, as

\begin{equation}\label{Error}
\Delta \chi^{2}(\eta) = 
\chi^{2}(\hat{\eta}+\delta \eta)
-\chi^{2}(\hat{\eta}) =
\sum_{i,j} \delta \eta_{i}  H_{ij}
\delta \eta_{j} \ ,
\end{equation}
where, $H_{ij}$  are the elements of the Hessian matrix which can be written as

\begin{equation}
H_{ij}=\frac{1}{2} \frac{\partial^{2} \chi^{2}}
{\partial \eta_{i} \partial \eta_{j}}
\bigg|_{\text{min}}  \,.
\end{equation}

The confidence region is normally can be given in the parameter space by supplying a value of $\Delta\chi^{2}$. 
In a standard parameter-fitting criterion, the errors are given by the choice of the tolerance ${\cal T} = \Delta\chi^{2}=1$ in Eq.~\eqref{Error}.
For the number of one fitted parameter ($N=1$), the confidence level (C.L.) is 68\% for $\Delta\chi^{2} =1$.
For the general cases and for more number of fitted parameters ($N > 1$), the $\Delta \chi^{2}$ should be calculated to determine the size of the resulting uncertainties.
The correct determination of the tolerance ${\cal T}$ should indicates that {\tt SKM18} fitting methodology as well as the uncertainties determination does correctly propagate the experimental uncertainty of $D^{*+}$ production data into the uncertainties of fitted charmed-meson FFs.

The determination of the size of uncertainties can be done by applying the ``Hessian
method'' based on the correspondence between the number of fitting parameters with the C.L. ${\cal P}$ and $\chi^{2}$.
The C.L. ${\cal P}$ can be written as follows

\begin{equation}
{\cal P} = \int_{0}^{\Delta \chi^{2}}\frac{1}{2\,
\Gamma(N/2)}
\left(\frac{\xi^{2}}{2}\right)^{\frac{N}{2} - 1}
e^{\left(-\frac{\xi^{2}}{2} \right)}
d\,\xi^{2} \,,
\end{equation}

where $\Gamma$ is the well-known Gamma function.
The value of $\Delta \chi^{2}$ is taken so that the C.L. becomes the one-$\sigma$-error range, namely ${\cal P}=0.68$. Similarly, for the 90$^{\rm th}$ percentile, one can use ${\cal P}=0.90$. Using the equation above, the value of $\Delta\chi^{2}$ is numerically calculated.

The Hessian matrix (or error matrix) is accessible by running the subroutine the CERN program library {\tt MINUIT}~\cite{James:1994vla}.
The uncertainty on a given observable ${\cal O}$, which is an attributive function of the input parameters of charmed-meson at the input scale $\mu_0^2$, is obtained by applying the ``Hessian method''.
Having at hand the value of $\Delta\chi^{2}$, and derivatives of the observables with respect to the charmed-meson fitted parameters, the ``Hessian method'' gives the uncertainties of a given observable ${\cal O}$. It reads, 

\begin{equation}
[\Delta {\cal O}_{i}]^{2}=\Delta \chi^{2} \sum_{j,k}
\left(\frac{\partial {\cal O}_{i} (\eta)}
{\partial \eta_j} \right)_{\hat \eta}
C_{j,k}	\left(\frac{\partial {\cal O}_{i}
(\eta)}{\partial \eta_{k}} \right)_{\hat \eta} \,,
\end{equation}

where $C_{j,k}$ is the inverse of the Hessian matrix, $C_{j,k} = H_{jk}^{-1}$.
For estimation of uncertainties at an arbitrary scale $\mu^{2}$, the obtained gradient terms are evolved by the DGLAP evolution kernel\cite{DGLAP},
and then the charmed-meson FF uncertainties as well as the uncertainties of other observables, such as total cross section for $e^+e^-$ annihilation into hadrons $\sigma_{tot}$, are calculated.

In the next section, we present the main results of this work, namely the ``{\tt SKM18}'' set of charmed-meson FFs at NLO and NNLO approximations.
First, we discuss the resultant $D^{*\pm}$-FFs and their uncertainties.
Then, we show the quality of the fits and compare the {\tt SKM18} theoretical predictions to the $D^{*\pm}$-meson production data sets.

%
\section{Discussion of SKM18 fit results} \label{sec:Results}
%

Now we turn to {\tt SKM18} numerical results for the global analysis of $D^{*\pm}$-FFs from SIA data. First, we present the {\tt SKM18} NLO and NNLO FFs for the charm and bottom densities in ZM-VFN scheme and investigate the dependence of obtained FFs on various data sets.
Second, we quantify the size of uncertainty bands due to the NNLO corrections. Next, {\tt SKM18} theoretical predictions will be compared to all SIA data  and other well-known phenomenological groups in literature.
We finalize this section with comments on the impact of collider data on  determination of gluon FFs and the reduction of FFs uncertainties.  Finally, in a detailed study we will consider the theoretical uncertainties due to the variation of the renormalization and factorization scales. Our study shows that the NNLO theoretical predictions  will indeed be more stable than the NLO ones. 

As was mentioned, the {\tt SKM18} includes for the fist time the NNLO  QCD corrections of  SIA cross section in order to achieve a high percentage accuracy. After extracting the NLO and NNLO $D^{*\pm}$-FFs, we will compare them with the NLO ones reported by KKKS08~\cite{Kneesch:2007ey} and the recent ones from AKSRV17~\cite{Anderle:2017cgl}. In addition, we will compare our theoretical predictions for the normalized cross sections with the analyzed experimental data.  
We use the public code {\tt APFEL}~\cite{Bertone:2013vaa} to compute the NLO and NNLO DGLAP evolution of the FFs and the SIA cross sections in the $\overline{\rm MS}$ factorization scheme in $z$-space. In this package the numerical solution for the time-like evolution equations are provided up to NNLO accuracy in QCD. Some of recent analyses on the FFs of light hadrons (pion, kaon and proton) have also been done by using this public code~\cite{Nocera:2017gbk,Bertone:2017tyb}.

Our results of the optimum fits for the charm and bottom FF parameters are presented in Tables.~\ref{tab:NLO} and \ref{tab:NNLO} for the NLO and NNLO fits, respectively. As was already mentioned, the parameter $N_c$ is basically unconstrained by data and comes with a relatively large error. For this reason, we calculate it from the first fit along with six free parameters and then we fix it in the second fit with five free parameters. 
In Tables.~\ref{Data-Set_NLO}, \ref{Data-Set_NNLO} the normalization shifts ${\cal N}_i$ for each data set and $\chi^2$ values are reported. We include 59 data points from single-inclusive hadron production in our global analysis of $D^{*\pm}$-FFs.
To see how much the NNLO approximation of $D^{*\pm}$-FFs improves the NLO ones, note that, for the NNLO global fit the value of $\chi^2$ is $1.27$ which is better than the NLO one, $\chi^2=1.31$. Normally, the quality of  fit  can be judged from the obtained fitted parameters and the $\chi^2$ values. According to the best fit parameters presented in Tables.~\ref{tab:NLO} and \ref{tab:NNLO}, one can conclude that our parameters are well determined.

%
\begin{table}[b]
\caption{ The individual $\chi^2$ values and the fitted normalization at NLO for each collaboration and the total $\chi^2$ fit for $D^{*\pm}$. More detailed discussion over the individual $D^{*\pm}$ data sets, and the definition of $\chi^2 (\{\xi_i\})$ are presented in the text. } 
\begin{ruledtabular}
\tabcolsep=0.06cm \footnotesize
\begin{tabular}{lccccc}
Collaboration& data & $\sqrt{s}$  GeV&  data   & ${\cal N}_i$  & $\chi^2$(NLO)\\
& properties &          &  points &        \\
&    &                  &                       \\\hline
ALEPH~\cite{Barate:1999bg}    & Inclusive  & 91.2 & 17 &  0.999006 &24.59  \\
& $b$-tagged  & 91.2 & 15 &  1.00104 &18.73  \\
OPAL~\cite{Ackerstaff:1997ki}  & Inclusive\ & 91.2 & 9 & 0.999305&2.02  \\
& $b$-tagged  & 91.2 & 9 &  0.999672 & 8.01  \\  
& $c$-tagged  & 91.2 & 9 &  1.002758 & 17.39\\\hline
{\bf TOTAL:} & & &59 &     &70.74                      \\ 
($\chi^{2}$/d.o.f) & & & & &1.31                        \\
\end{tabular} \label{Data-Set_NLO}
\end{ruledtabular}
\end{table}
%

%
\begin{table}[b]
\caption{As in Table.~\ref{Data-Set_NLO} but in NNLO QCD accuracy.}
\begin{ruledtabular}
\tabcolsep=0.06cm \footnotesize
\begin{tabular}{lccccc}
Collaboration& data & $\sqrt{s}$  GeV&  data   & ${\cal N}_i$  & $\chi^2$(NNLO)        \\
& properties &          &  points &                                \\
&    &                  &                                          \\ \hline
ALEPH~\cite{Barate:1999bg}    & Inclusive  & 91.2 & 17 & 0.998900&24.51          \\
& $b$-tagged   & 91.2 & 15 &  1.000990 &17.99                     \\
OPAL~\cite{Ackerstaff:1997ki}  & Inclusive \ & 91.2 & 9 & 0.999099&1.92             \\
& $b$-tagged   & 91.2 & 9 &  0.999700 &7.61    \\
& $c$-tagged   & 91.2 & 9 &  1.002699 &16.94    \\ \hline
{\bf TOTAL:} & & &59 &     &68.97                           \\
($\chi^{2}$/d.o.f) & & & & &1.27                          \\
\end{tabular}\label{Data-Set_NNLO}
\end{ruledtabular}
\end{table}
%

%
\begin{table*}[t]
\caption{\label{tab:NLO} Fit parameters for the fragmentation of b- and c-quark into the $D^{*\pm}$-meson at NLO.
The starting scale is taken to be $\mu_{0} ^2= 18.5$~GeV$^2$ for charm and  bottom quark. The values labeled by (*) have been fixed after the first minimization, since the available SIA data dose not constrain all unknown fit parameters well enough.  }
\begin{ruledtabular}
\begin{tabular}{cccccc}
flavor $i$ &$N_i$ & $\alpha_i$ & $\beta_i$                             \\ \hline
$c, \overline{c}$ &$67.031^*$ & $1.908\pm0.0194$&$1.133\pm0.070$        \\
$b, \overline{b}$ & $5.742\pm1.574$&$ 0.994\pm0.0385$&$3.249\pm0.279$   \\
\end{tabular}
\end{ruledtabular}
\end{table*}

%
\begin{table*}[t]
\caption{\label{tab:NNLO} Fit parameters for the bottom and charm FFs into the  charged $D^{*\pm}$-meson at NNLO.
The starting scales and other explanations are as in Table.~\ref{tab:NLO}. } 
\begin{ruledtabular}
\begin{tabular}{cccccc}
flavor $i$ &$N_i$ & $\alpha_i$ & $\beta_i$                             \\ \hline
$c, \overline{c}$ &$53.896^*$& $1.854\pm0.0191$&$1.170\pm0.069$        \\
$b, \overline{b}$ & $5.127\pm1.351$&$ 0.967\pm0.0372$&$3.248\pm0.274$  \\
\end{tabular}
\end{ruledtabular}
\end{table*}

%
\subsection{ {\tt SKM18} FFs and comparison with other FF sets } 
%

\begin{figure*}[htb]
\begin{center}
\vspace{0.50cm}
\resizebox{0.48\textwidth}{!}{\includegraphics{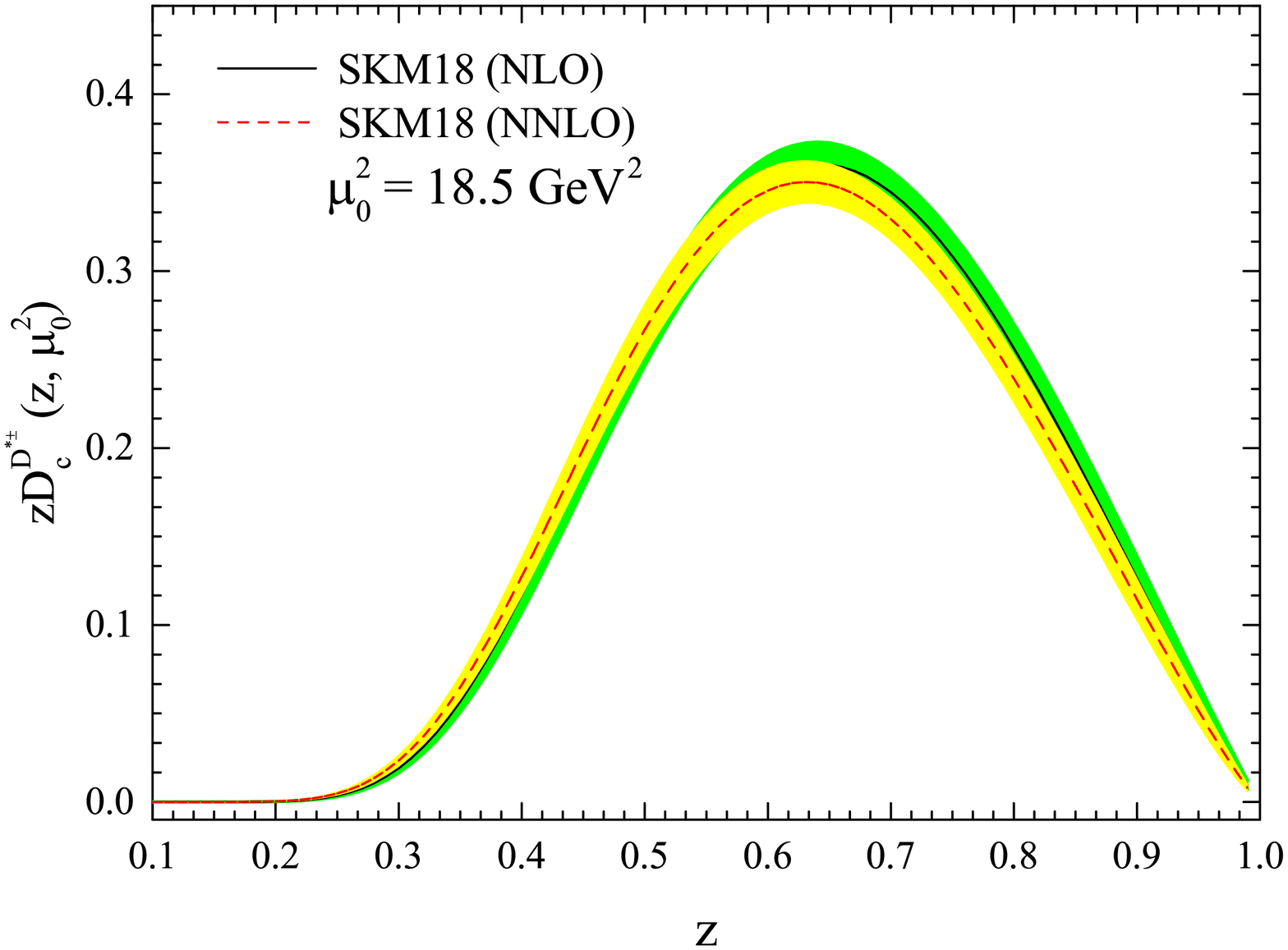}}  
\resizebox{0.48\textwidth}{!}{\includegraphics{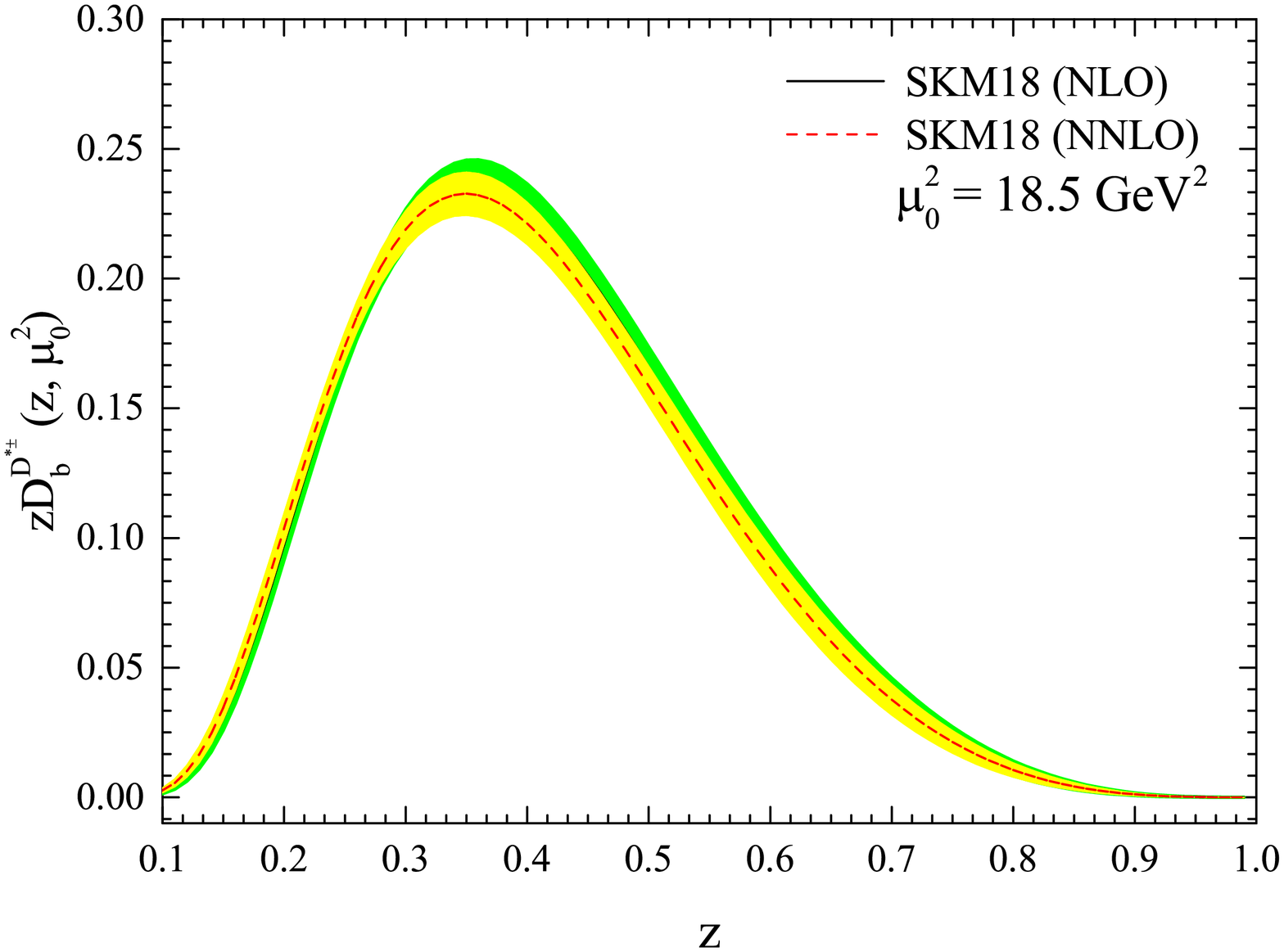}} 
\caption{ {\tt SKM18} fragmentation densities and their uncertainties (shaded bands) are shown for $zD^{D^{*\pm}}_i$ at the initial scale $\mu_0^2=18.5$~GeV$^2$ for $c $ and  $b$ both at NLO (solid lines) and NNLO (dashed lines). } \label{fig:FFsQ0}
\end{center}
\end{figure*}

\begin{figure*}[htb]
	\begin{center}
		\vspace{0.50cm}
		\resizebox{0.48\textwidth}{!}{\includegraphics{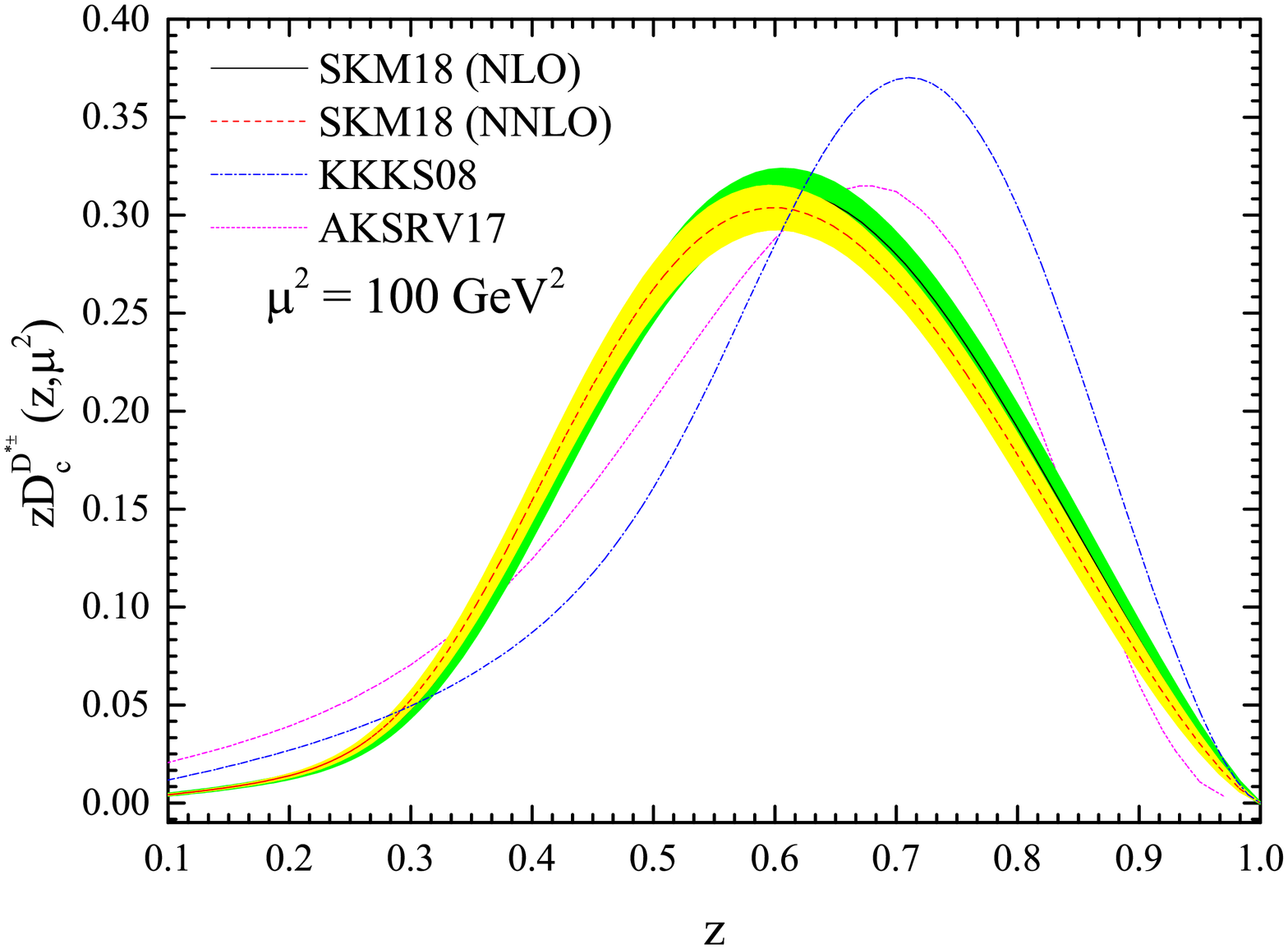}} 
		\resizebox{0.48\textwidth}{!}{\includegraphics{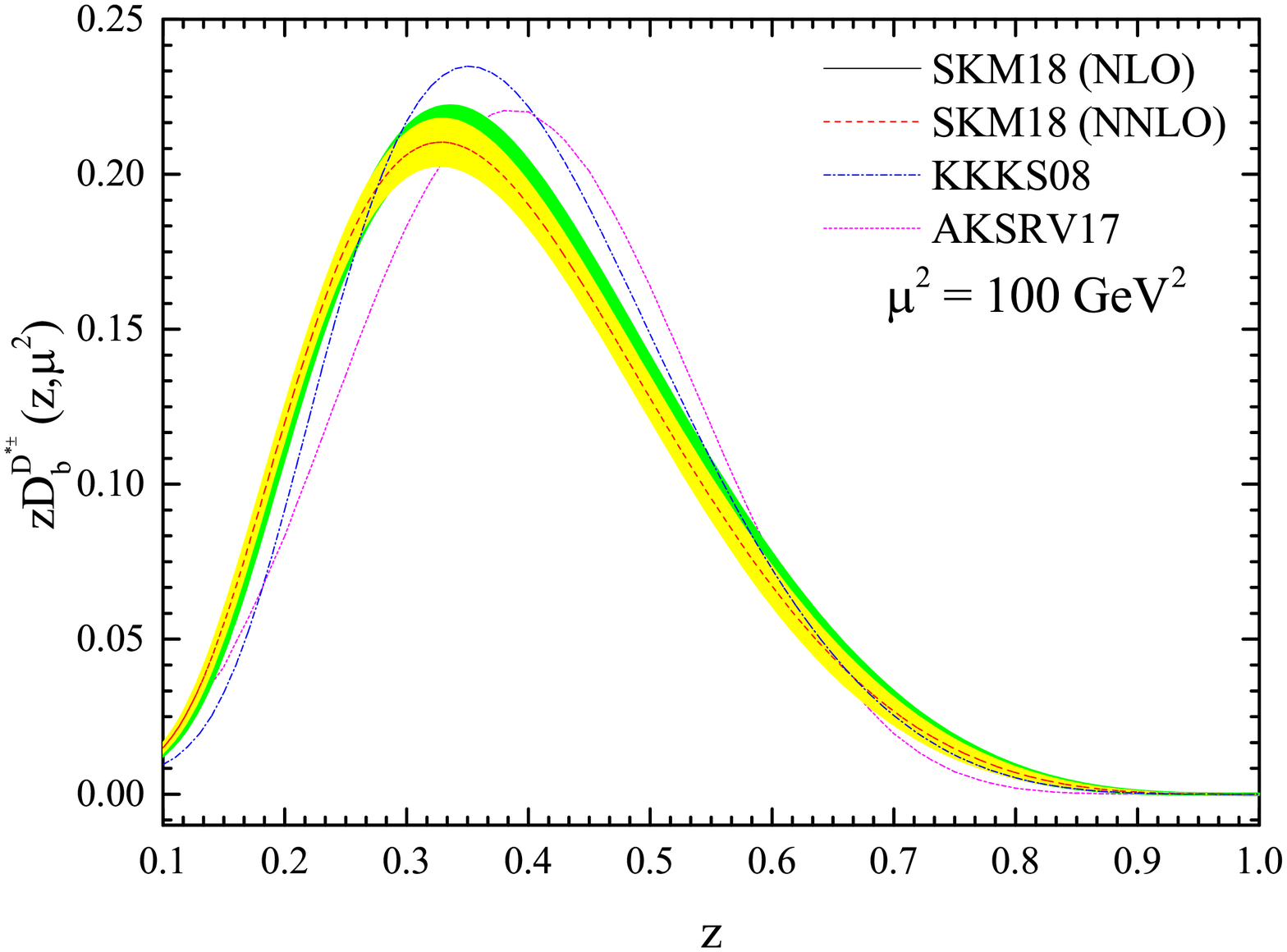}} 		
		\resizebox{0.48\textwidth}{!}{\includegraphics{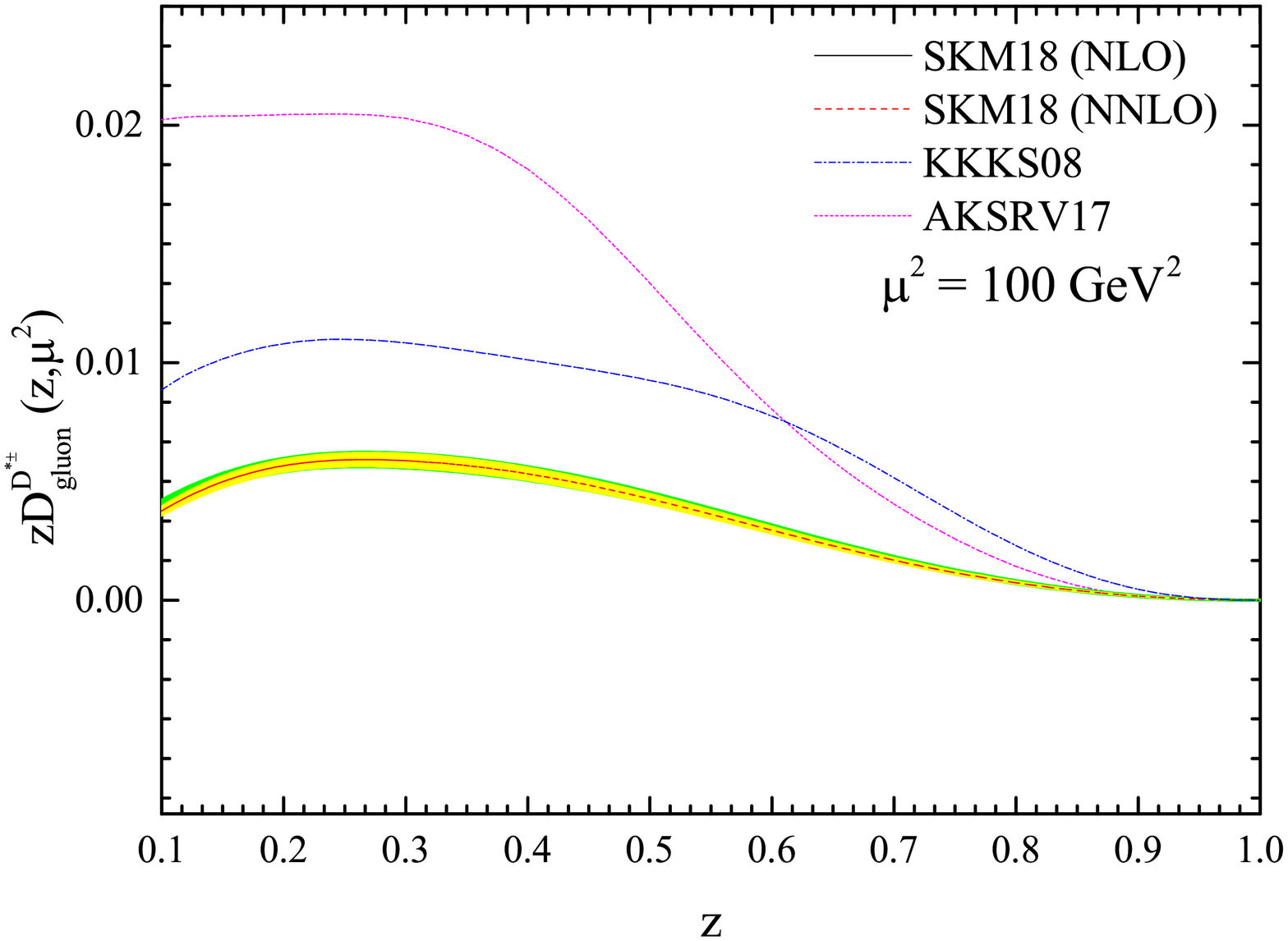}} 		 
		\caption{ Fragmentation densities and their uncertainties (shaded bands) are shown for $zD^{D^{*
		\pm}}_i$ at  $\mu ^2 =100$~GeV$^2$ for $c$, $b$ and gluon both at NLO (solid lines) and NNLO (dashed lines). Our results are also compared with the KKKS08 (dot-dashed lines) \cite{Kneesch:2007ey} and the AKSRV17 (short dashed lines)~\cite{Anderle:2017cgl} results  at NLO. } \label{fig:FFsQ}
	\end{center}
\end{figure*}
\begin{figure*}[htb]
	\begin{center}
		\vspace{0.50cm}
		\resizebox{0.48\textwidth}{!}{\includegraphics{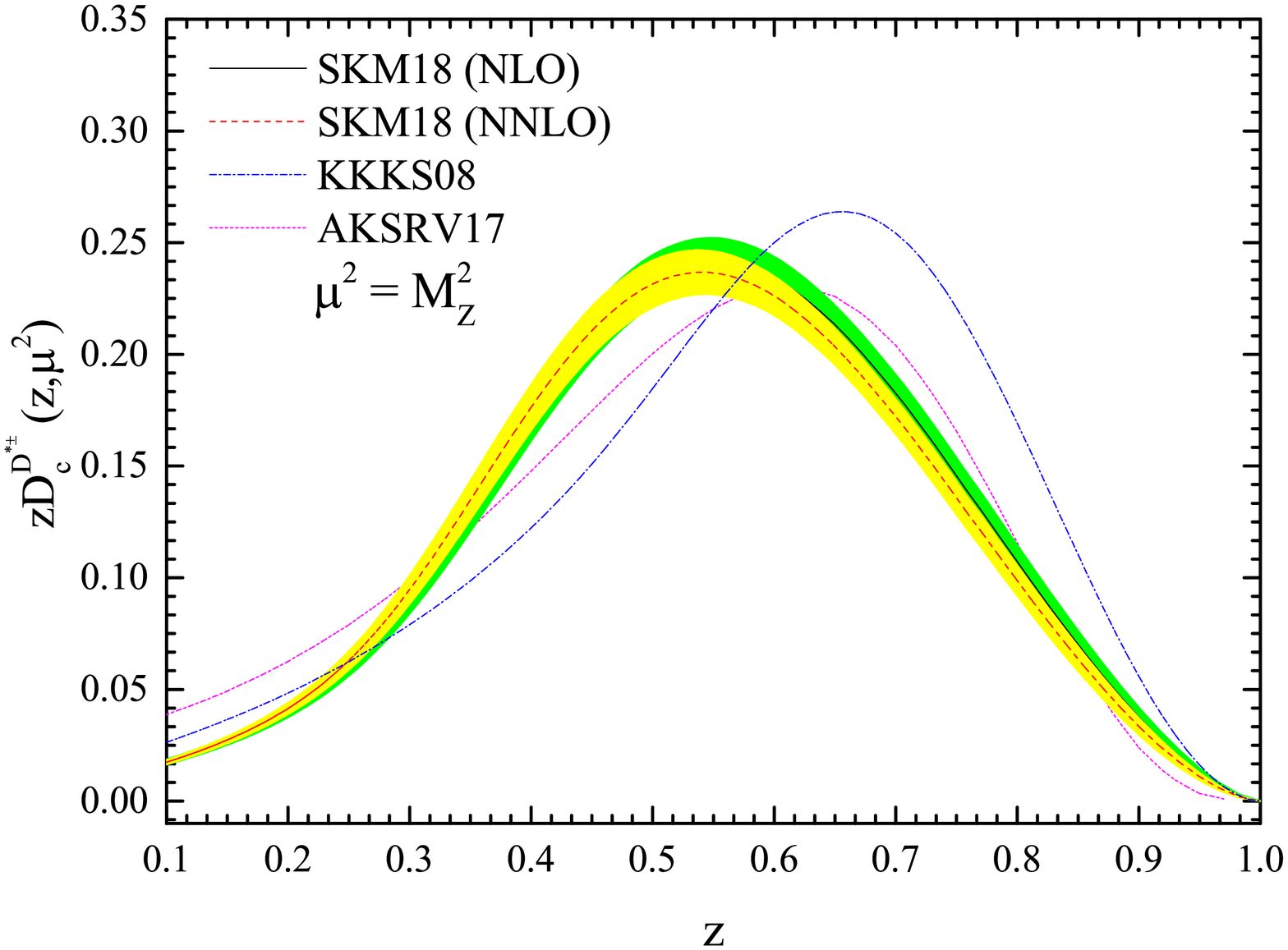}} 
		\resizebox{0.48\textwidth}{!}{\includegraphics{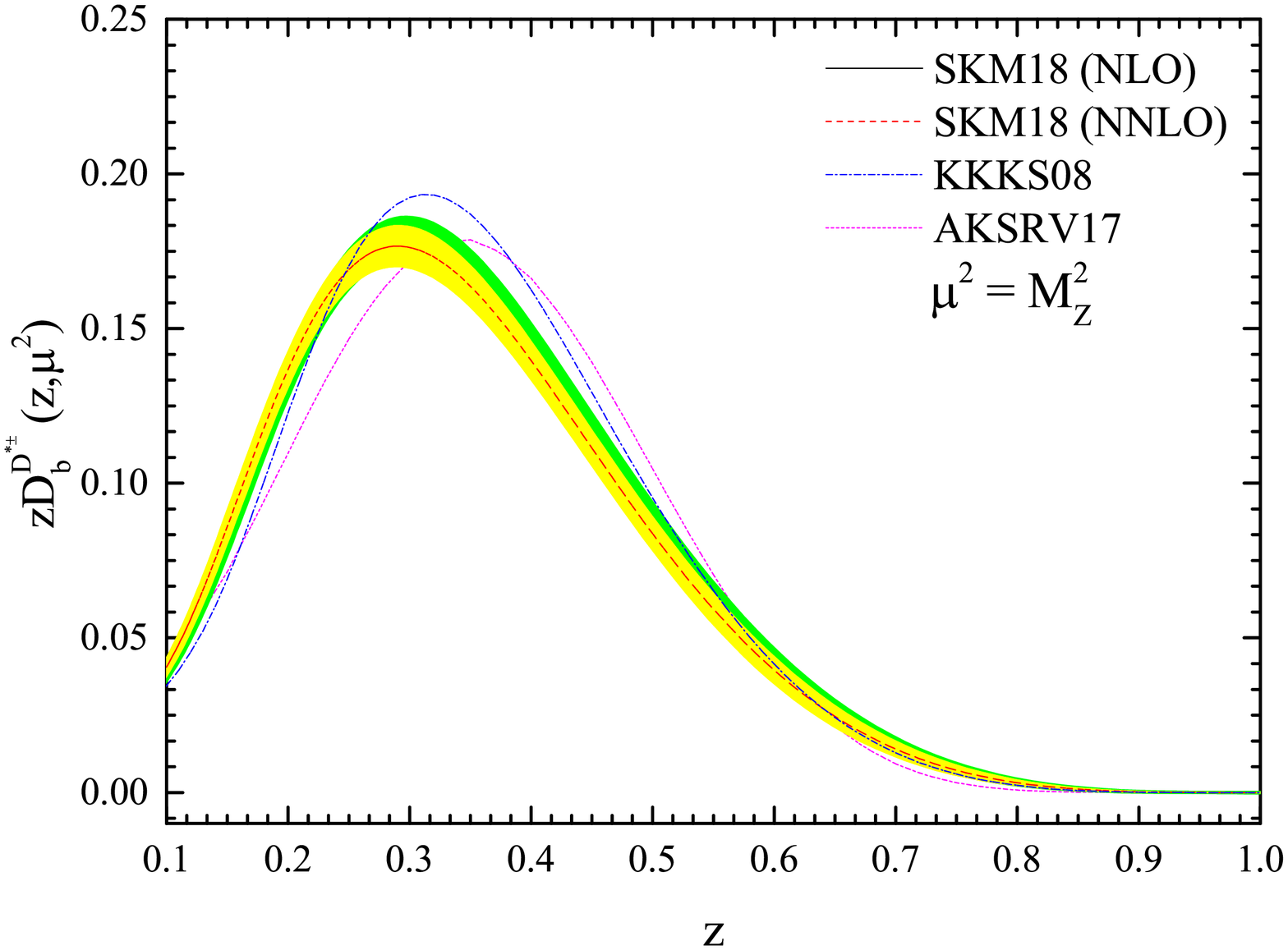}} 		
		\resizebox{0.48\textwidth}{!}{\includegraphics{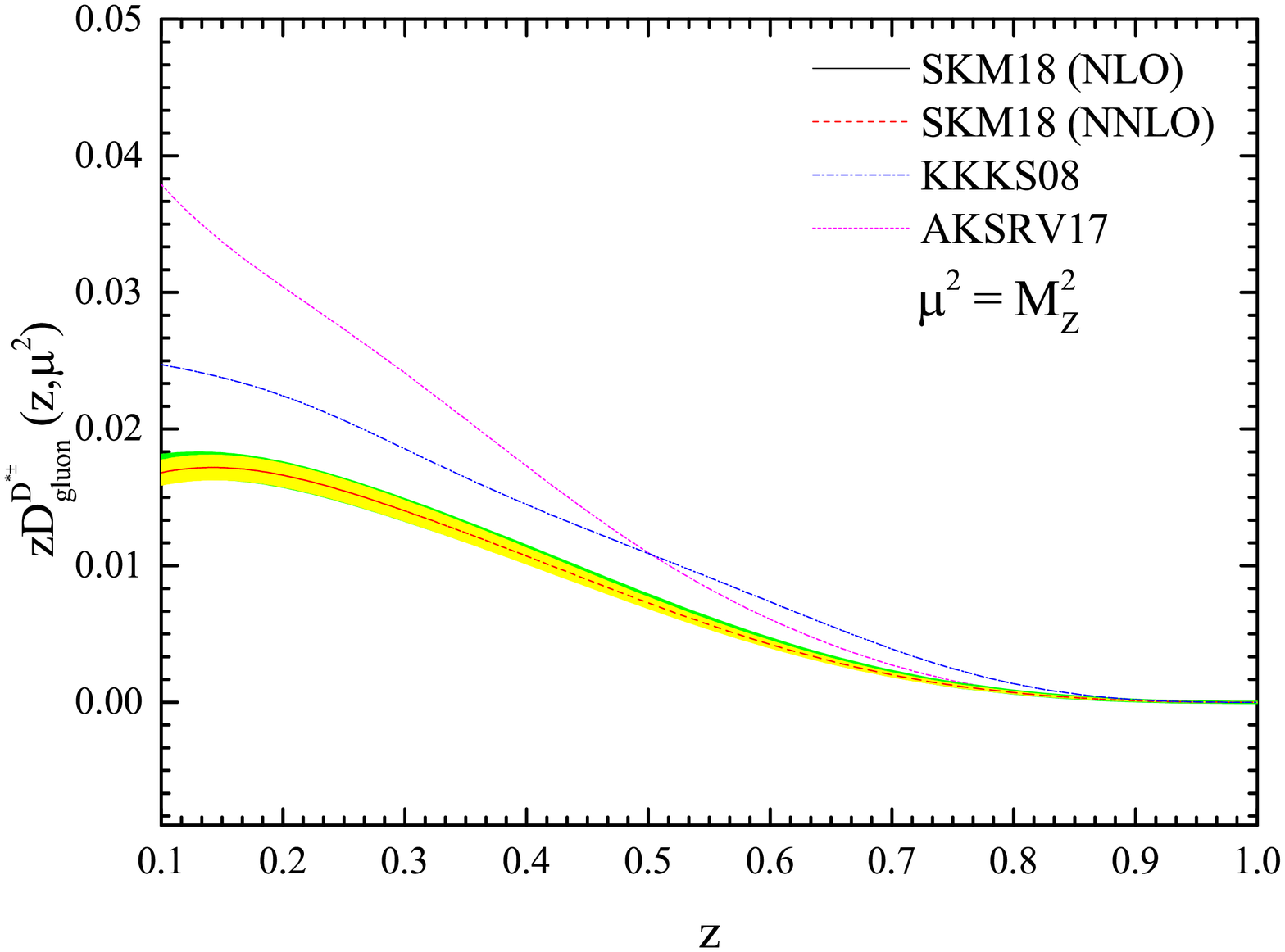}}   
		\caption{Fragmentation densities and their uncertainties (shaded bands) are shown for $zD^{D^{*\pm}}_i$ at  $\mu ^2 =M_Z^2$ for $c$, $b$ and gluon at  NLO (solid lines) and NNLO (dashed lines).  Our results are also compared with the KKKS08 (dot-dashed lines) \cite{Kneesch:2007ey} and the AKSRV17 (short dashed lines)~\cite{Anderle:2017cgl} results  at NLO.} \label{fig:FFsMZ}
	\end{center}
\end{figure*}

The {\tt SKM18} $D^{*\pm}$ fragmentation densities and their uncertainties are presented in Figs.~\ref{fig:FFsQ0}, \ref{fig:FFsQ} and \ref{fig:FFsMZ}.
The results of this analysis include the one-$\sigma$ uncertainty bands. They are compared to the central value from the KKKS08 analysis~\cite{Kneesch:2007ey} and very recent analysis of AKSRV17~\cite{Anderle:2017cgl}.
Now, we take this opportunity to discuss the {\tt SKM18} $D^{*\pm}$-FFs  in more detail. In Fig.~\ref{fig:FFsQ0}, the $c$ and $b$ FFs including their uncertainty bands are shown at NLO (solid lines) and NNLO (dashed lines) at the initial scale $\mu_0^{2}$, in which we consider $\mu _0^2 =18.5$~GeV$^2$ both for the charm and bottom quark FFs. A sightly small difference is evident between the NLO and NNLO results.  A basic question is now that, what impact the higher order QCD corrections can have on the reduction of FFs uncertainties, specifically the NNLO corrections  in comparison with the NLO ones? To this end, we show in Fig.~\ref{fig:FFsQ0} the {\tt SKM18} NNLO FFs with their error bands. From this figure one can conclude that the NLO and NNLO uncertainties are slightly similar in size. 

The resulting {\tt SKM18} FFs are depicted in Fig.~\ref{fig:FFsQ} for higher values $\mu^2 > \mu_0^2$. The  NLO (solid lines) and NNLO (dashed lines) results have been shown for the $c$- and $b$-quarks and also for the gluon at $\mu ^2 =100$~GeV$^2$.
In this figure, we compared our results with the NLO ones from KKKS08~\cite{Kneesch:2007ey} (dot-dashed lines)~\cite{KKKS08code} and with very recent fit of AKSRV17 (short dashed lines)~\cite{Anderle:2017cgl}.  
In comparison to the AKSRV17 analysis, the {\tt SKM18} charm-quark  FF is similar to the one from AKSRV17  while the contribution of gluon FF extracted in AKSRV17 is significantly larger than the {\tt SKM18} and  KKKS08 one. A main reason for this difference is due to the fact  that the AKSRV17 allows for a nonzero gluon FF at the initial scale due to inclusion of hadron-hadron collision data in their fit. We will show that a similar behavior is observed for the charm and gluon FFs at $\mu ^2=M_Z^2$. Obviously, the {\tt SKM18} b-quark FF behaves similar to the one from the AKSRV17 and KKKS08 analysis, although, the result of KKKS08 for the c-quark FF is not compatible with the one from {\tt SKM18} and AKSRV17. 

In Fig.~\ref{fig:FFsMZ}, we repeat the same study as before so the $c $, $b$ and gluon fragmentation densities are shown at $\mu ^2=M_Z^2$ at NLO (solid lines) and NNLO (dashed lines). They are also compared with the KKKS08 (dot-dashed lines) and the AKSRV17 analyses (short dashed lines). As one can conclude from Fig.~\ref{fig:FFsMZ}, the most difference between our results and the KKKS08 and the AKSRV17 analysis is in the gluon FF, as was already discussed its reason. While the bottom FF in the analysis of {\tt SKM18}, AKSRV17 and KKKS08 are in agreement, the charm one behaves differently. One of the differences between the {\tt SKM18} and AKSRV17 analysis and the KKKS08 analysis is the charm tagged cross section $D^{*\pm}$ from OPAL collaboration which has not been included in the  KKKS08 fit. The charm FF is constrained by the charm-tagged data  which is included both in the {\tt SKM18} and AKSRV17 analysis.

\begin{figure*}[htb]
	\begin{center}
		\vspace{0.50cm}
		\resizebox{0.48\textwidth}{!}{\includegraphics{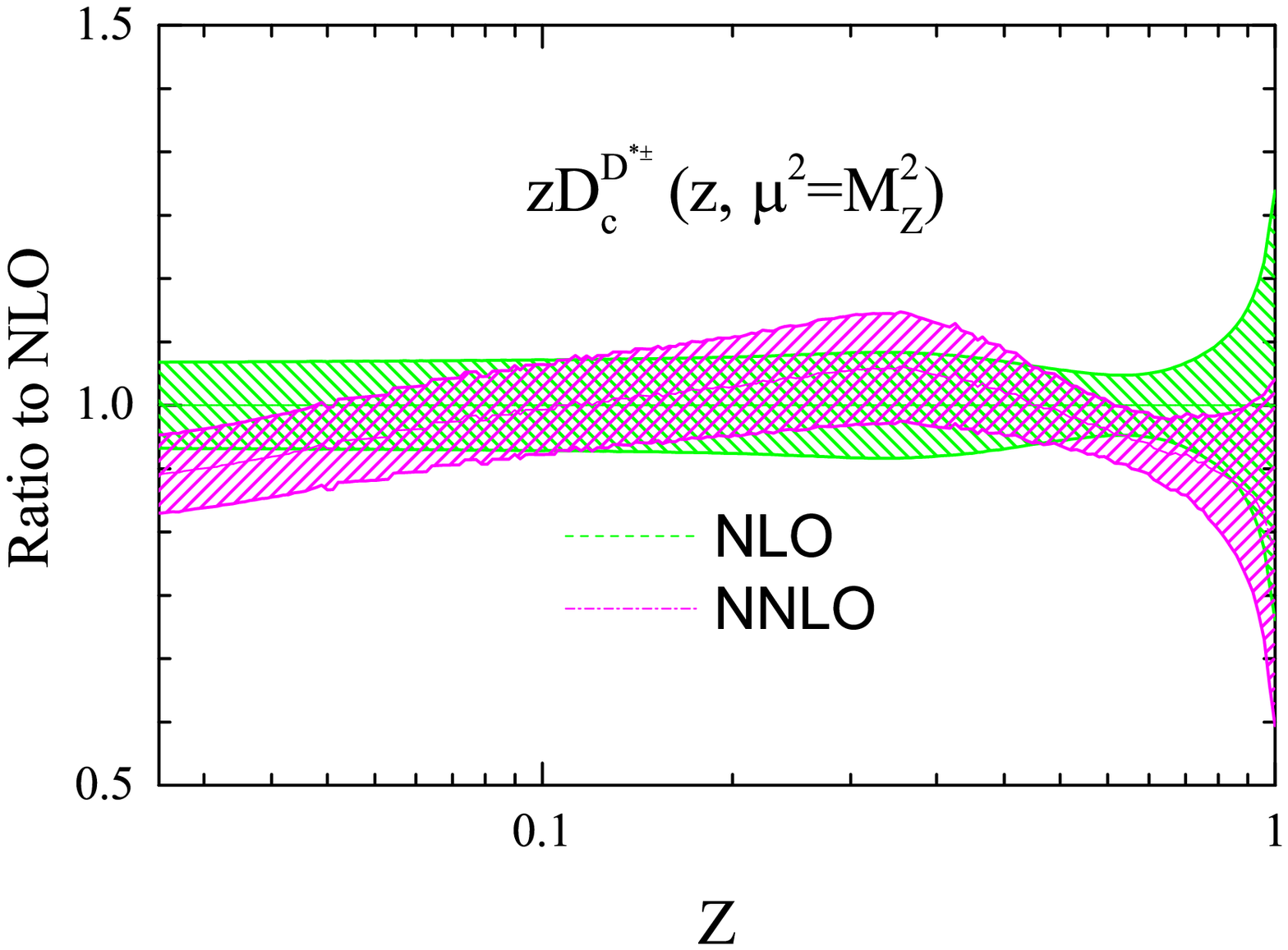}}  
		\resizebox{0.48\textwidth}{!}{\includegraphics{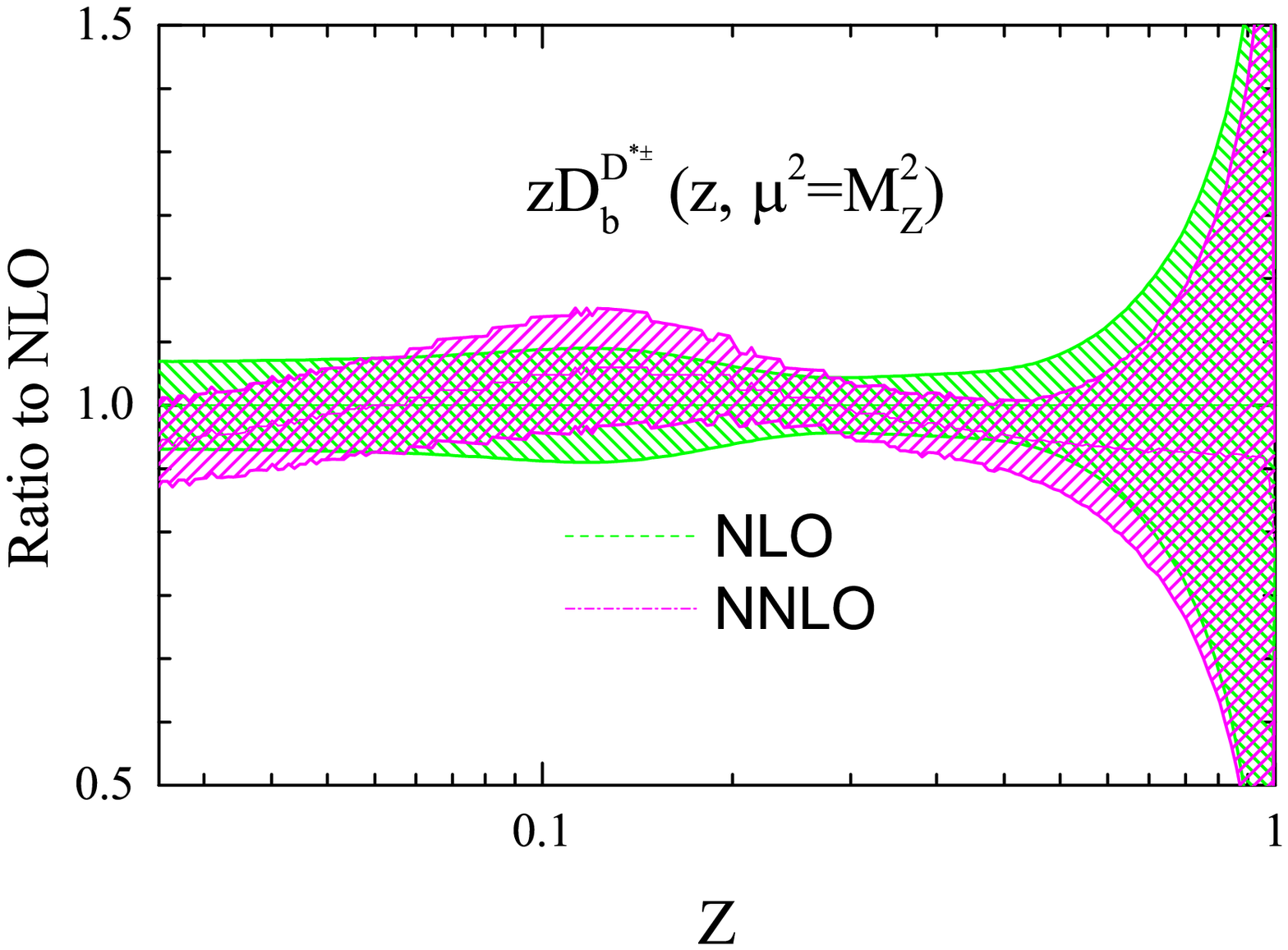}}  
		\resizebox{0.48\textwidth}{!}{\includegraphics{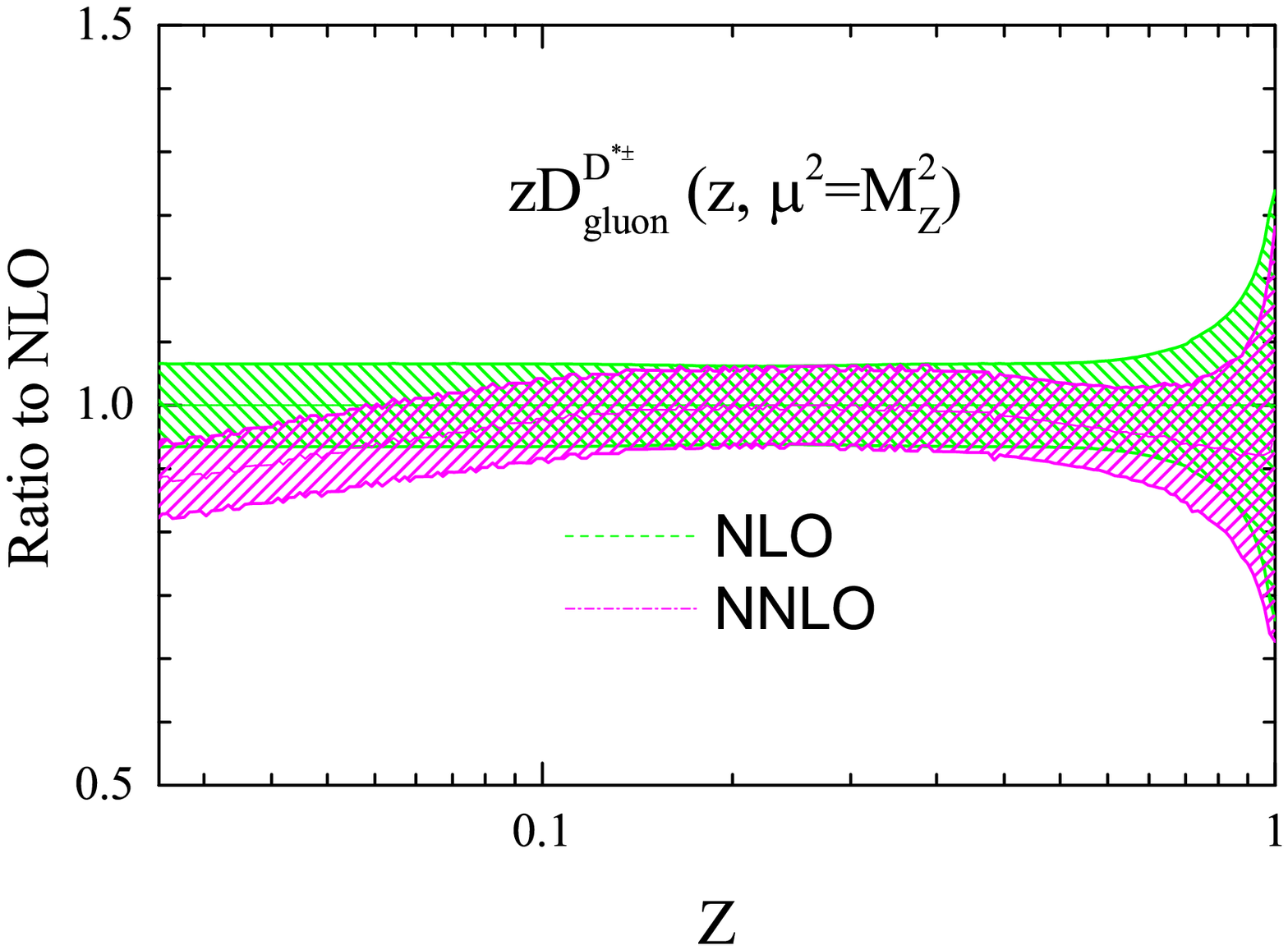}}  
		\caption{ The NNLO/NLO ratios for {\tt SKM18} fragmentation densities and their uncertainties are shown for $zD^{D^{*\pm}}_i$ at  $\mu ^2 =M_Z^2$ for $c$, $b$ and gluon at NLO (dashed lines) and NNLO (dot-dashed lines).} \label{fig:RaioFFsMZ}
	\end{center}
\end{figure*}

Let us now discuss on the resulting FFs and their uncertainties by focusing on their perturbative convergence as well as the effects arising due to the inclusion of higher-order QCD corrections. Concerning the {\tt SKM18} fit quality of the total dataset, the most noticeable feature is the improvement upon inclusion of higher-order corrections. Although, the improvement of the $\chi^2/{\rm n.d.f}$ is  rather marginal when going from NLO to NNLO.  This finding demonstrates that the inclusion of the NNLO QCD corrections slightly improves the description of the data. A further noticeable aspect of the comparisons in Figs.~\ref{fig:FFsQ0}, \ref{fig:FFsQ} and \ref{fig:FFsMZ} is related to the size of the {\tt SKM18} FFs uncertainties which show that the NLO and NNLO uncertainties are slightly similar in the size. One can also conclude from the presented plots that the differences between the NLO and NNLO FFs are slightly small. We should stress here that, from Ref.~\cite{Bertone:2017tyb} the same pattern for the {\tt NNPDF} FFs can be seen  at NLO and NNLO.
As the {\tt NNPDF} collaboration are mentioned in their recent paper, while in some cases the LO and NLO distributions can  differ  by more than one standard deviation, the differences between the NLO and NNLO FFs are relatively small so the uncertainty sizes of the NLO and NNLO FFs  are similar, as well. 
For more qualitative studies of the size of reductions on the FF uncertainties  due to inclusion of higher-order QCD corrections,  in Fig.~\ref{fig:RaioFFsMZ} we present the NNLO/NLO ratio for the FF of  $c$, $b$ and gluon at $\mu ^2 =M_Z^2$. From Fig.~\ref{fig:RaioFFsMZ} one can now judge the reduction of error band widths  as well as the size of improvements upon inclusion of higher-order corrections.

As was mentioned in Sec.~\ref{sec:parametrization}, at the initial scale $\mu_0$ the FFs of gluon and light quarks ($q=u, d, s$) are set to zero and they are evolved to higher scales using the DGLAP  equations. The light quark contributions to produce the $D^{*\pm}$-mesons are very small in comparison with the heavy quarks so one expects them  to have main contributions in production of light hadrons such as pion, kaon and proton. For this reason, in Figs.~\ref{fig:FFsQ} and \ref{fig:FFsMZ} we restricted our results to the charm, bottom and gluon FFs. As can be seen from these figures, the differences between the {\tt SKM18}, AKSRV17  and KKKS08 analysis  are considerable for the charm and gluon FFs compared to the bottom quark one.
Generally, this is most likely due to the different experimental data which are included in these analysis. Individually, we list some possible explanations for this finding.  Firstly, as we mentioned in Sec.~\ref{sec:data selection}, the KKKS08 fit~\cite{Kneesch:2007ey} have included both the CLEO and withdrawn Belle data at the low energy $Q=10.5$~GeV.
Since, these data are bellow the bottom threshold, they affect on the charm and anti-charm FFs,  directly. On the other hand, because of using CLEO and Belle data, the KKKS08 analysis have included the mass effects of $D^{*\pm}$-meson and heavy quarks into their analysis, according to Eqs.~\eqref{xp} and \eqref{xp+cross}, which might be the source of some differences. 
Secondly, charm-tagged cross section $D^{*\pm}$ from OPAL collaboration which has not been included in the KKKS08 fit are used both in the analysis of {\tt SKM18} and AKSRV17.
Finally, the AKSRV17 collaboration have used the data from hadron-hadron collisions and jet fragmentation in the proton-proton scattering in addition to the electron-positron data. 
The collider data is directly sensitive to the gluon FFs. The collider data is taken into account in the AKSRV17 analysis so could carry  a large amount of information on the gluon FFs, and may also account for the differences observed between these results.
The $b$-FFs plotted in Figs.~\ref{fig:FFsQ} and \ref{fig:FFsMZ} are quite similar to the ones obtained by the KKKS08 and AKSRV17 analysis, because in both fits the bottom-tagged data from ALEPH and OPAL~\cite{Barate:1999bg,Ackerstaff:1997ki} are included and the bottom FFs are significantly constrained by these data sets. 

It should be pointed  out that, the authors of AKSRV17~\cite{Anderle:2017cgl} analysis claimed who  applied the online FFs generator~\cite{KKKS08code} to extract the KKKS08 FFs. We are particularly puzzled since there is inconsistency between their KKKS08 curves and the results which can be obtained by the FF generator. We do not judge in this respect. 

\begin{figure}[htb]
\begin{center}
\vspace{0.50cm}
\resizebox{0.550\textwidth}{!}{\includegraphics{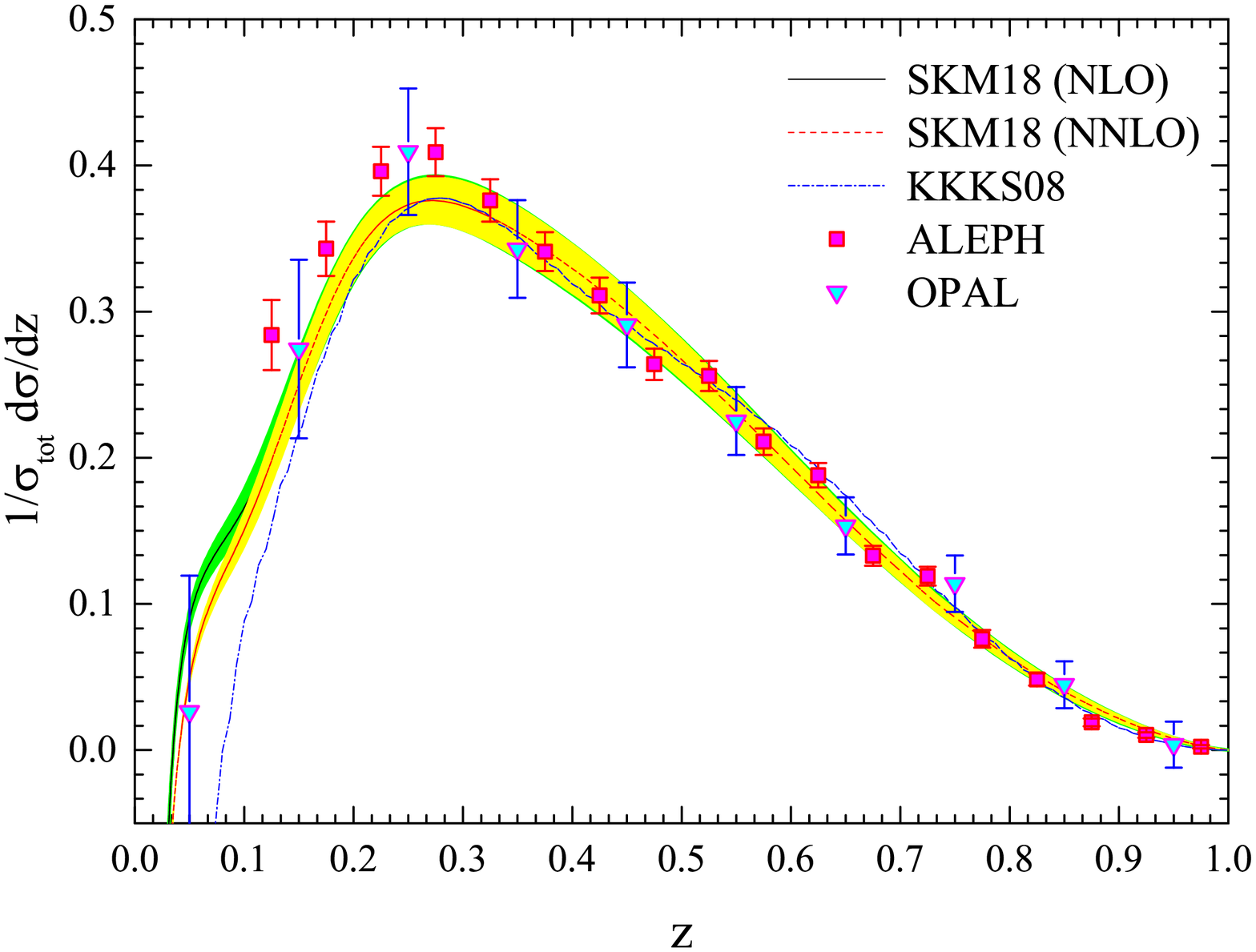}}  
\caption{ Our NLO (solid line) and NNLO (dashed line) results for the normalized total cross sections of $D^{*\pm}$-production compared with the KKKS08 ones~\cite{Kneesch:2007ey} (dot-dashed line) at the scale $Q=M_Z$. Data from ALEPH~\cite{Barate:1999bg} and OPAL~\cite{Ackerstaff:1997ki} are also shown in this scale. The shaded bands refer to our uncertainty results at NLO (green band) and NNLO (yellow band).} \label{fig:total}
\end{center}
\end{figure}
\begin{figure}[htb]
\begin{center}
\vspace{0.50cm}
\resizebox{0.550\textwidth}{!}{\includegraphics{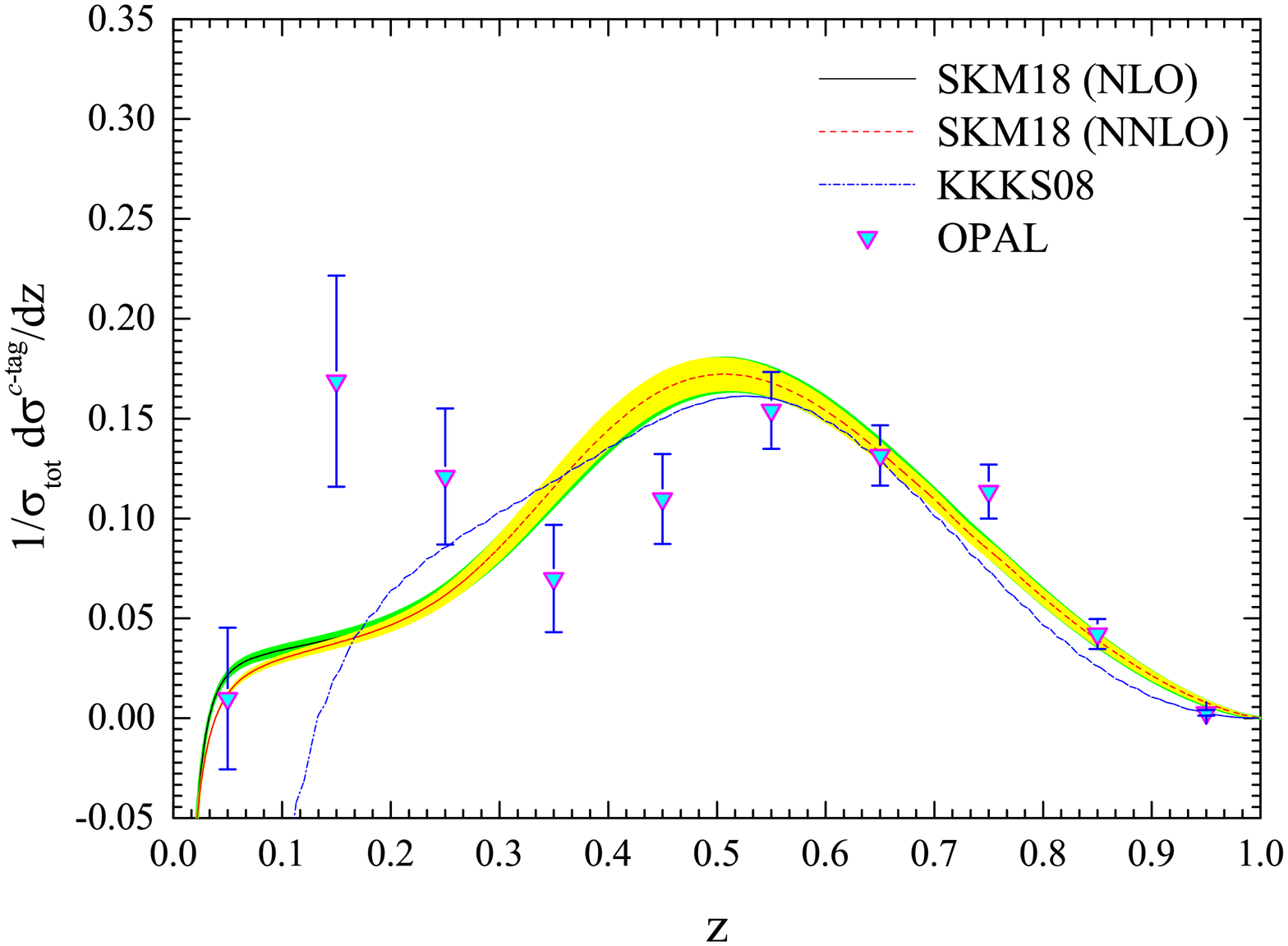}}  
\resizebox{0.550\textwidth}{!}{\includegraphics{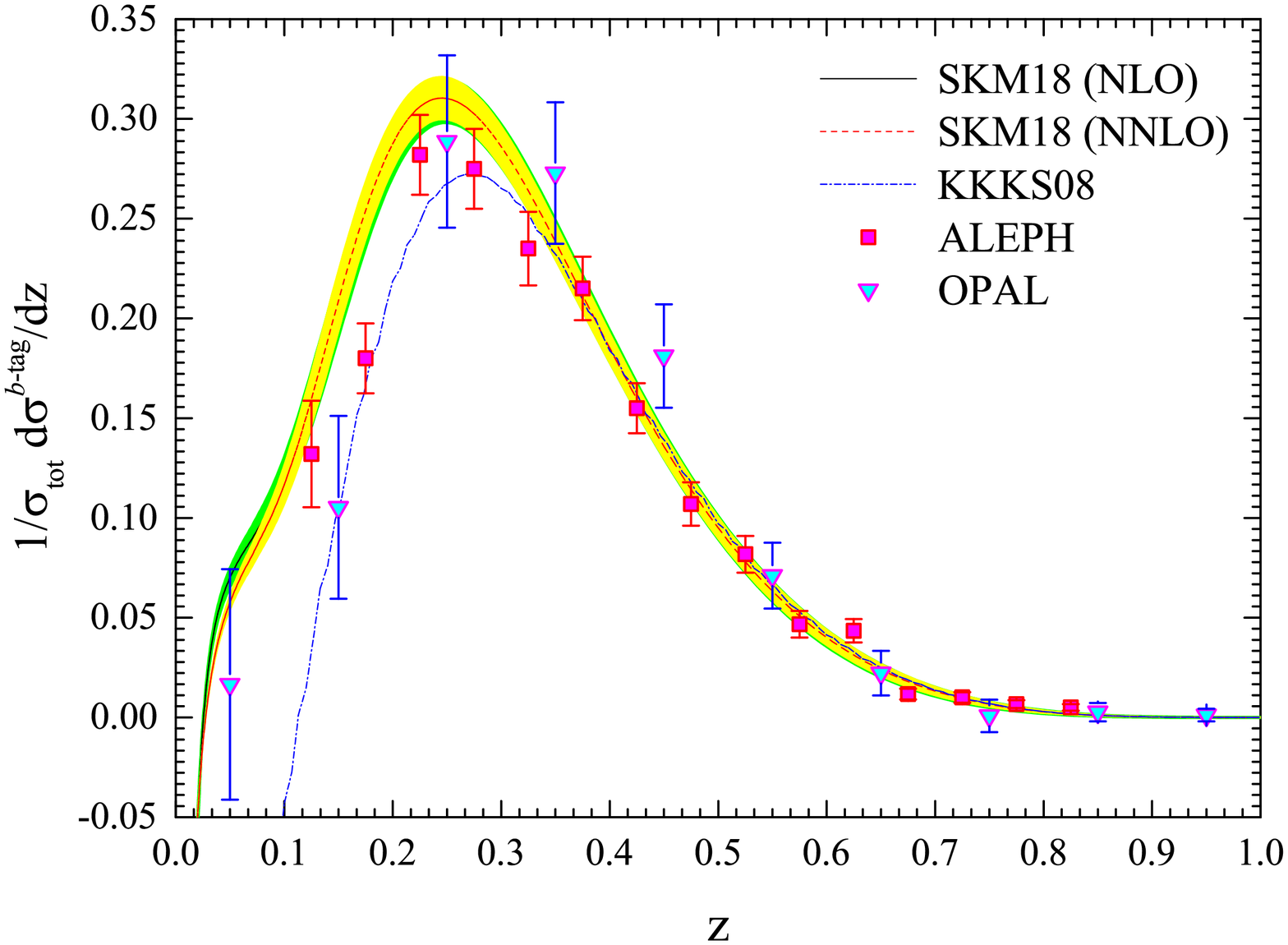}}  
\caption{ Our NLO (solid line) and NNLO (dashed line) results for the normalized charm and bottom tagged  cross sections of $D^{*\pm}$-production compared with the KKKS08 ones~\cite{Kneesch:2007ey} (dot-dashed line) at the scale $Q=M_Z$. Data  from ALEPH~\cite{Barate:1999bg} and OPAL~\cite{Ackerstaff:1997ki} are also shown in this scale.
The shaded bands refer to our uncertainty results at NLO (green band) and NNLO (yellow band).} 
\label{fig:cbtag}
\end{center}
\end{figure}

%
\subsection{ Discussion of fit quality and data/theory comparison} 
%

After our detailed discussion on the {\tt SKM18} FFs in comparison to the results in literature, we now turn to present our theoretical predictions for the SIA cross sections.
In Figs.~\ref{fig:total} and \ref{fig:cbtag}, the theoretical calculations based on the {\tt SKM18} global QCD analysis are compared to the analyzed available data.
The ALEPH~\cite{Barate:1999bg} and OPAL~\cite{Ackerstaff:1997ki} normalized total cross sections at $\mu^2 = M_Z^2$ are shown in Fig.~\ref{fig:total} along with our calculations for the SIA cross section at NLO (solid line) and NNLO (dashed line) accuracies. According to these results we find that our NLO and NNLO QCD fits are in good agreement with the experimental data and our theoretical predictions describe the data well for the intermediate and large values of $z$.

Towards small-$z$ region, our model fall down the data  because the evolution equation become unstable in small-$z$ region. The splitting functions  lead to negative FFs for $z<<1$ in evolution equations, and additionally mass corrections play important role in this region. So we exclude regions where mass corrections and the singular small-$z$ behavior of the splitting functions are effective. As we mentioned in Sec.~\ref{sec:data selection}, we exclude the regions with $z<0.1$ and $z>0.95$ in our global fit. Also we compare our theoretical results with the KKKS08 model in Figs.~\ref{fig:total} and \ref{fig:cbtag}. The AKSRV17 theoretical predictions are not shown because they are not available to the authors. According to~\cite{Kneesch:2007ey}, the KKKS08 analyses are done at both ZM-VFN  and  GM-VFN schemes. On the other hand, they calculate the free parameters from Belle/CLEO (at $10.5$~GeV), ALEPH/OPAL (at $M_Z$) and global data fits separately. Since our analyses base on the ZM-VFN scheme and we only include the ALEPH and OPAL data, then we compare our results with the corresponding KKKS08 set of FFs in the ZM-VFNS which are calculated with ALEPH/OPAL data fit. As one can see in Figs.~\ref{fig:total} and \ref{fig:cbtag}, the KKKS08 set is not compatible with data at the small $z$ region.
In Fig.~\ref{fig:cbtag} the ALEPH~\cite{Barate:1999bg} and OPAL~\cite{Ackerstaff:1997ki} normalized charm and bottom tagged cross sections at $\mu^2 = M_Z^2$ are shown and also our fit results are presented for this cross section at NLO (solid line) and NNLO (dashed line). 
The most visible conclusion is the very considerable agreements of {\tt SKM18} with the analyzed data. From Fig.~\ref{fig:total} one can conclude that the KKKS08 overestimate the data for the small values of $z$. This pattern is also seen in Fig.~\ref{fig:cbtag}.  In the next section, we aim to present our predictions for the energy spectrum of $D^{*\pm}$-mesons produced in polarized and unpolarized top quark decays.

%
\subsection{Importance of theoretical uncertainty at NNLO} 
%

The possible sources of uncertainties on the FFs classify into the experimental errors on the data, and the theoretical or phenomenological assumptions in the global QCD fit. The theoretical uncertainties include, for example, higher order QCD effects in the calculation of cross sections, the assumption of flavor or charge conjugation symmetries, the parametrization forms of the FFs at an arbitrary initial scale, and so on. 

One source of uncertainties on the theoretical results is the values selected for the renormalization ($\mu_R$) and factorization ($\mu_F$) scales, so the former associated with the renormalization of the strong coupling constant. In principle, one can use two different values for these scales, however, a choice often made consists of setting $\mu_F=\mu_R=\mu$ and we adopted this convention for the results presented.

\begin{figure*}[htb]
\begin{center}
\vspace{0.50cm}
\resizebox{0.48\textwidth}{!}{\includegraphics{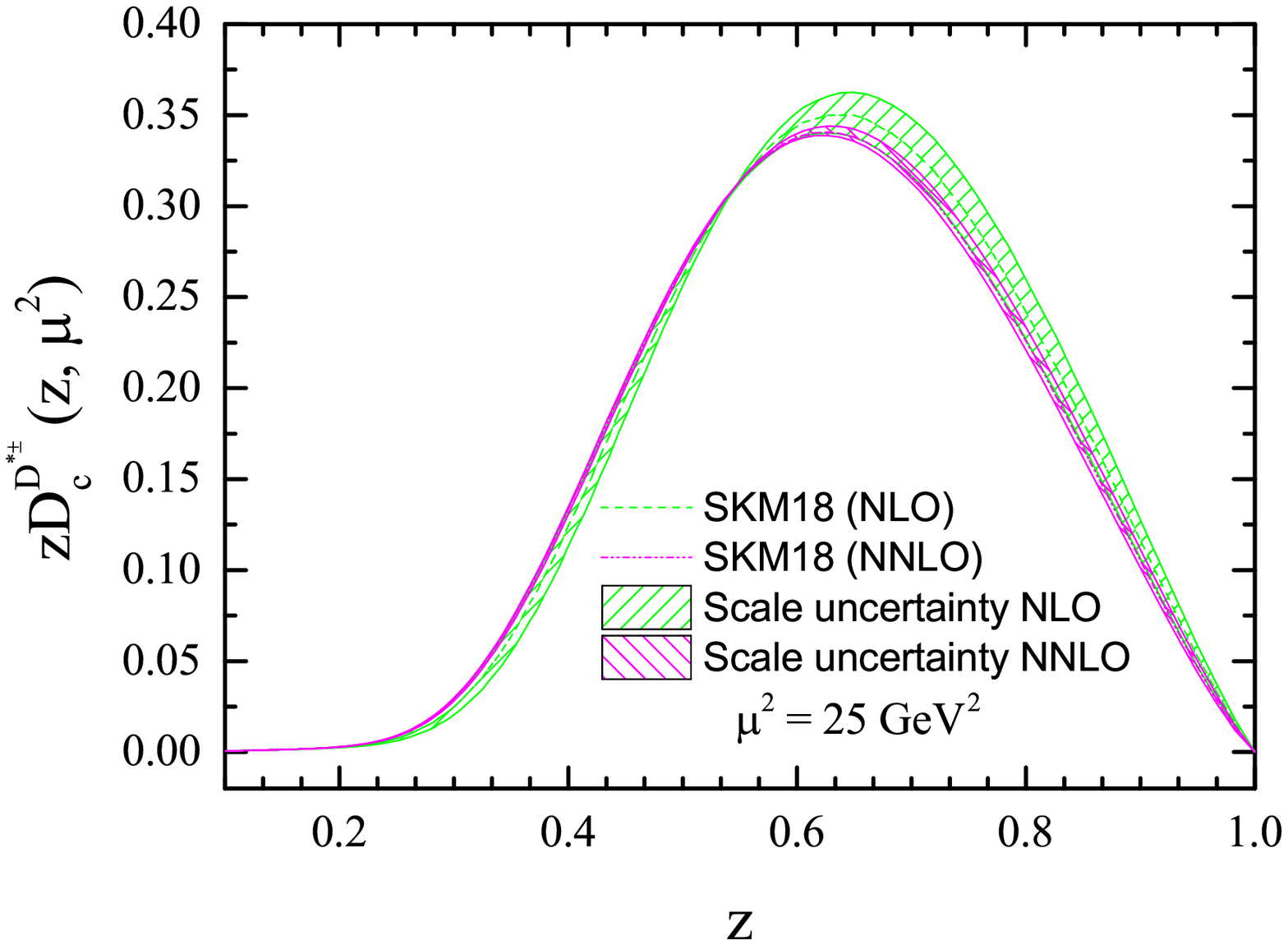}} 
\resizebox{0.48\textwidth}{!}{\includegraphics{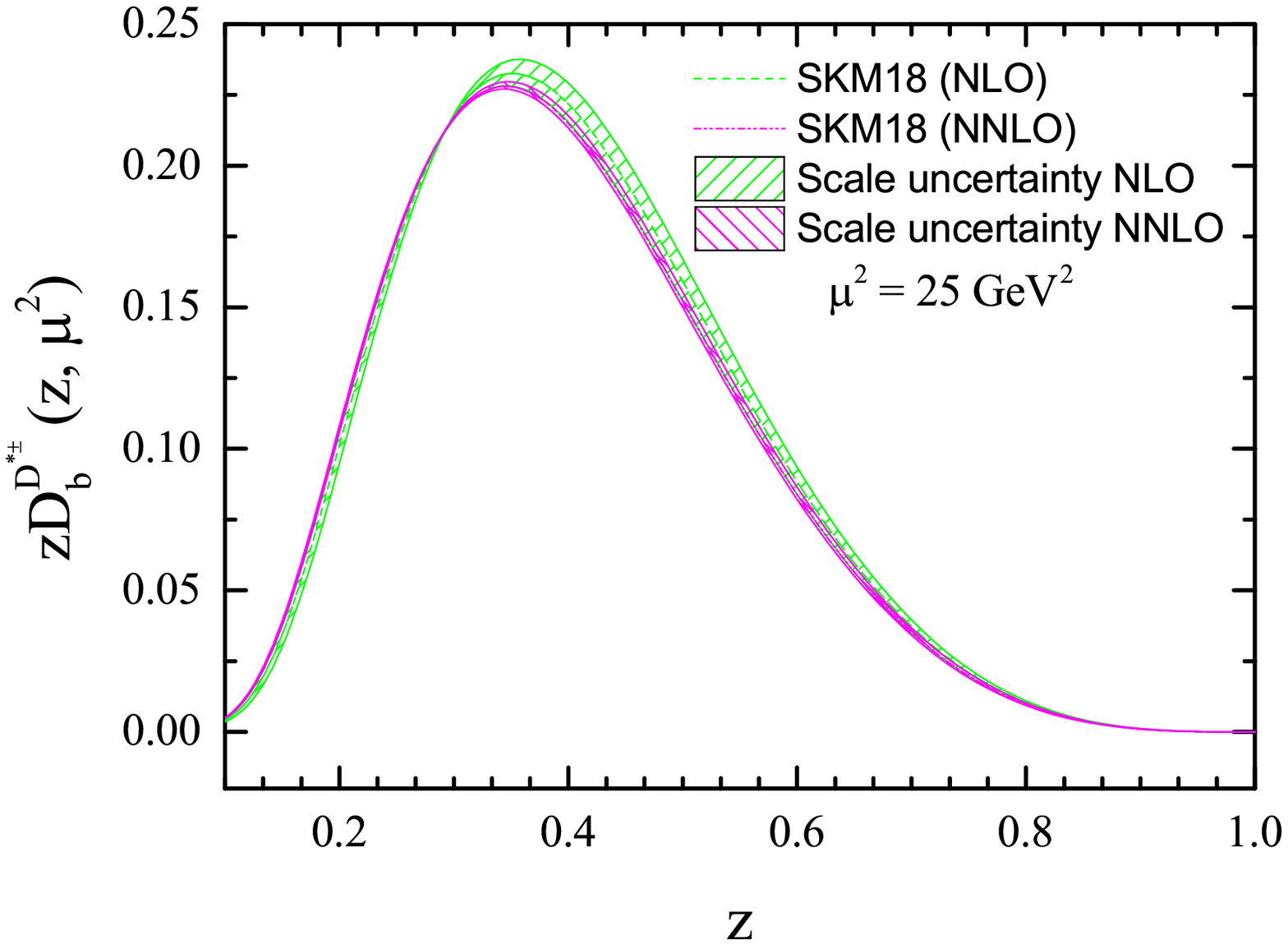}} 		
\resizebox{0.48\textwidth}{!}{\includegraphics{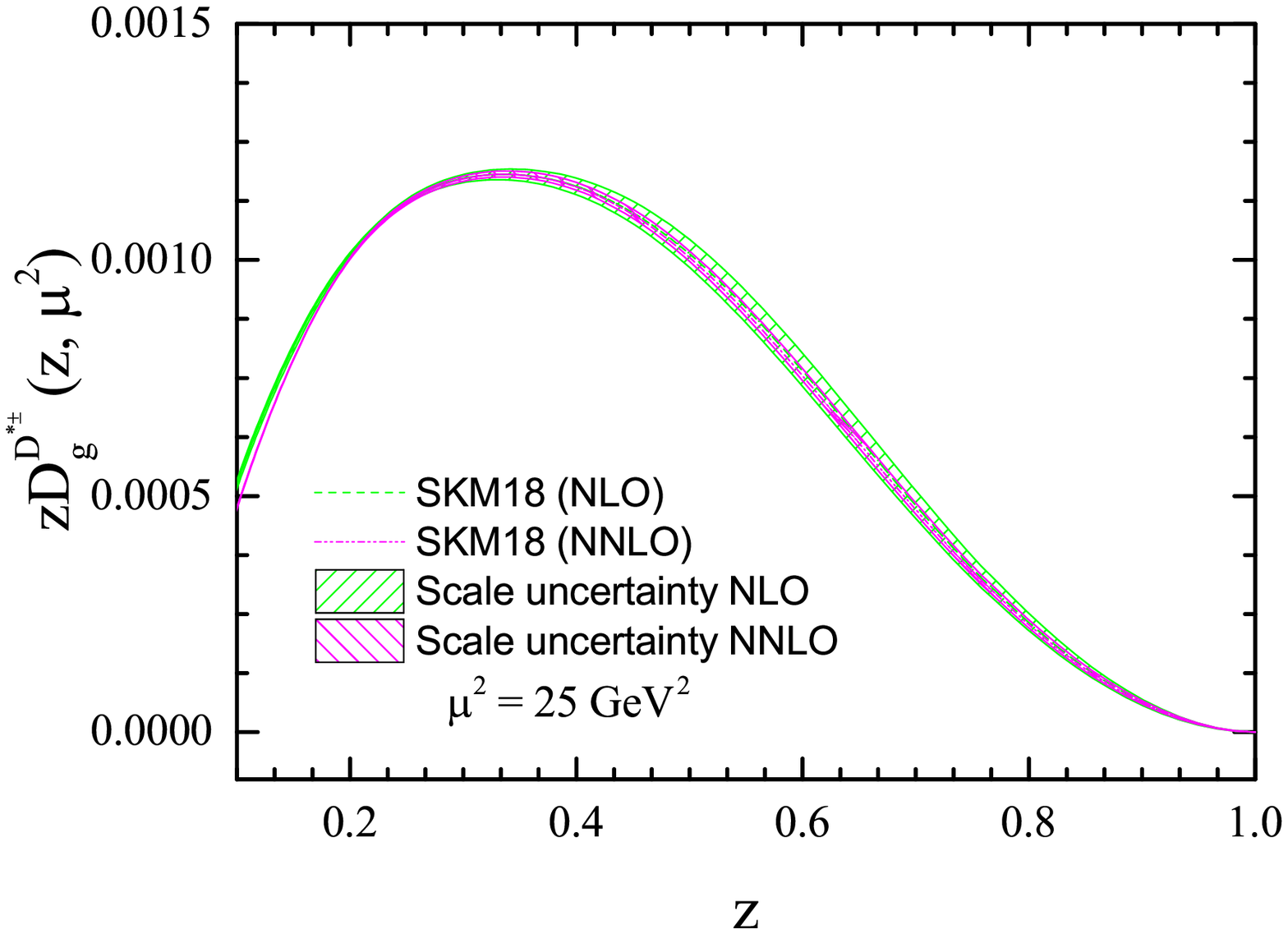}} 		 
\caption{ Fragmentation densities $zD^{D^{*
\pm}}_i(z, \mu^2)$ are shown  at  $\mu^2=Q^2=25$~GeV$^2$ for the gluon and the charm and bottom quarks both at NLO (dashed lines) and NNLO (dot-dashed lines). As a common convention, here we set $\mu_F=\mu_R=\mu$. The shaded bands indicate theoretical uncertainties when the scale is ranged as $Q/2 \le\mu \le 2Q$.} \label{fig:FFsRen}
\end{center}
\end{figure*}

In this section, we will examine the importance of these scales  at the NLO and NNLO approximations and the variation effects of these scales  on the theoretical results. \\
The scale dependence of FFs is shown in Fig.~\ref{fig:FFsRen}, where the results of NLO and NNLO accuracy (shaded bands)  are shown for $Q/2 \le\mu \le 2Q$. As is seen, the theoretical calculations depend on the choice of the scale $\mu$ and it is, indeed, one of the most important source of the theoretical uncertainties. We expect that this kind of theoretical uncertainties shrinks progressively when  higher order corrections are  included. Therefore, the NNLO predictions will indeed be more stable than the NLO ones. This is exactly what one can conclude from the {\tt SKM18} results presented in Fig.\ref{fig:FFsRen}. 

%
\section{Energy spectrum of inclusive $D^{*\pm}$-mesons in polarized and unpolarized top decays} \label{sec:top decay}
%

In this section, we turn to apply the extracted $D^{*\pm}$-FFs to make our phenomenological predictions for the energy spectrum of charmed mesons produced in top quark decays through the following process
\begin{eqnarray}\label{pros}
t \rightarrow b + W^+ (g) \rightarrow W^+ D^{*\pm} + X,
\end{eqnarray}
where $X$ stands for the unobserved final state. The study of energy distributions of produced mesons through top quark decays might be introduced as an indirect channel to search for the top quark properties. In this channel, both the $b$-quark and the gluon may hadronize into the charmed-meson so the gluon contributes to the real radiations at NLO and higher orders. 
Note that, the contribution of gluon FF can not be discriminated so that this contribution is being calculated to see where it contributes to the energy spectrum of produced meson. Therefore, this part of calculation is of more theoretical relevance, however in the scaled-energy of hadrons, as an experimental quantity, all contributions including the b-quark and gluon contribute.

\begin{figure}
	\begin{center}
		\includegraphics[width=0.7\linewidth,bb=137 42 690 690]{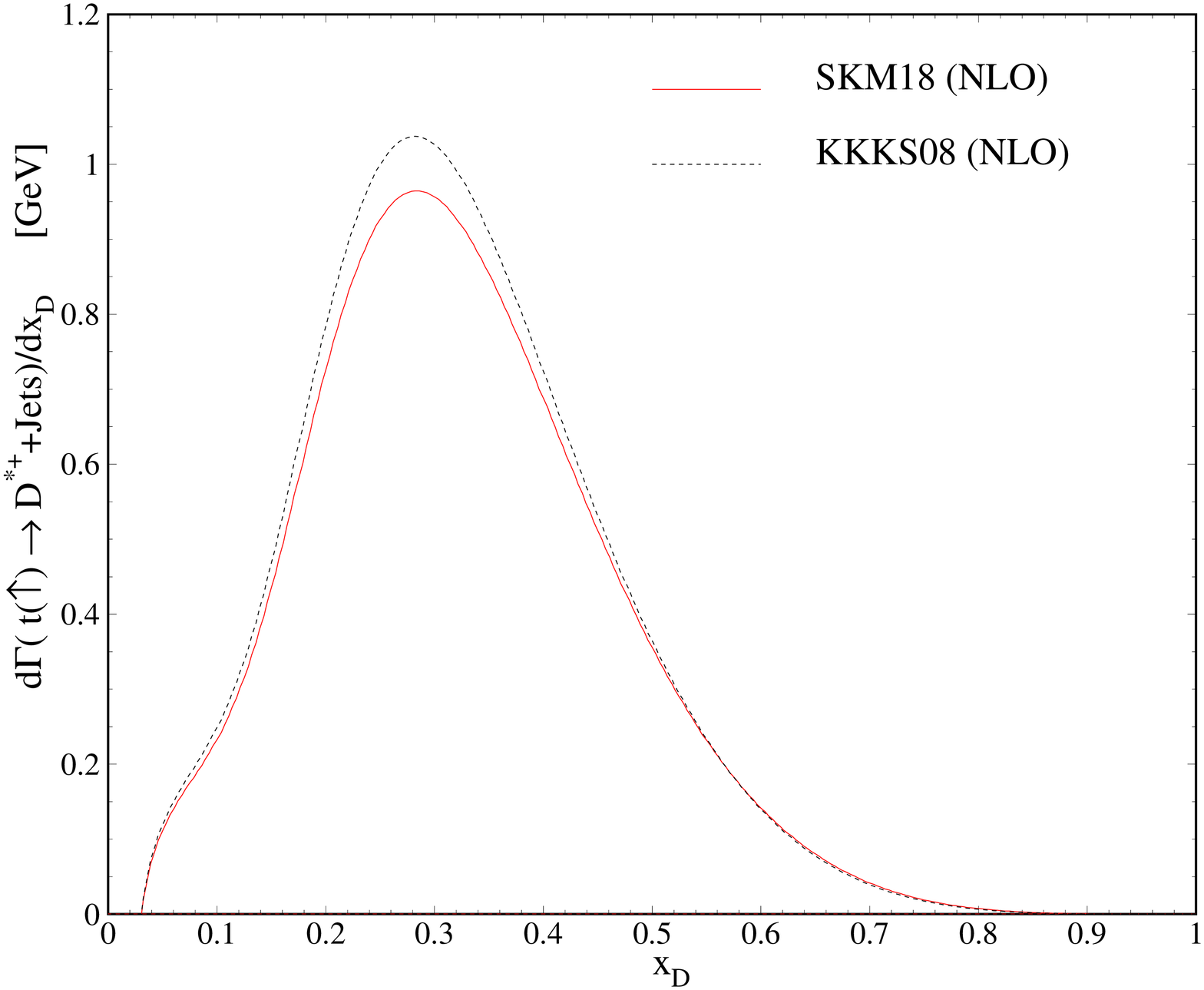}
		\caption{\label{fig1}%
			$x_D$ distribution of $d\Gamma^{\textbf{NLO}}/dx_D$ for the polarized top quark decay  at $\mu_F=m_t$. We also show the same distribution using the
			FFs presented by KKKS08  \cite{Kneesch:2007ey} at NLO.}
	\end{center}
\end{figure}

\begin{figure}
	\begin{center}
		\includegraphics[width=0.7\linewidth,bb=137 42 690 690]{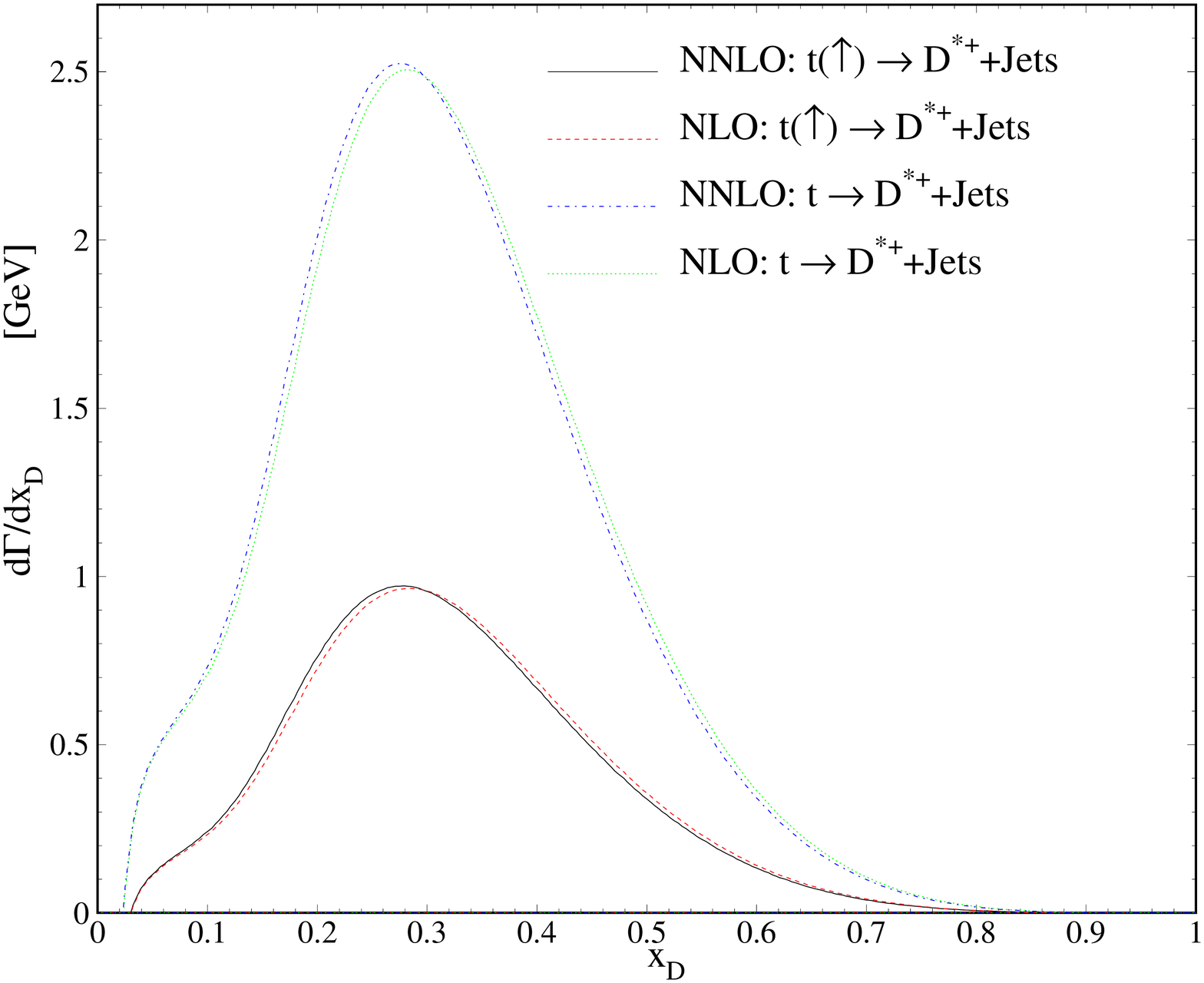}
		\caption{\label{fig2}%
			$d\Gamma(t\to D^{*\pm}+Jets)/dx_D$ as a function of $x_D$ at NLO for the polarized (dashed) and unpolarized (dots) top decays. We also show the same results at NNLO.}
	\end{center}
\end{figure}

Ignoring the $b$-quark mass and by working in the ZM-VFN scheme, to obtain the energy distribution of $D^{*\pm}$-meson in the process (\ref{pros}), we employ the factorization theorem~\eqref{convolution} as  
\begin{eqnarray}\label{eq:master}
\frac{d\Gamma}{dx_D}=\sum_{i=b,g}\int_{x_i^{min}}^{x_i^{max}}
\frac{dx_i}{x_i}\,\frac{d\Gamma}{dx_i}(\mu_R,\mu_F)
D_i^{D^{+\pm}}\left(\frac{x_D}{x_i},\mu_F\right), \nonumber \\
\end{eqnarray}
where, following Ref.~\cite{Kniehl:2012mn}, we define the scaled-energy fraction of $D^{*\pm}$ as $x_D=2E_{D}/(m_t^{2}-m_W^{2})$  and
$d\Gamma/dx_i $ are the parton-level differential decay rates of the process
$t\to i+W^+ (i=b,g)$.
The  analytical expressions for the differential decay widths $d\Gamma/dx_i$ at the QCD NLO approximation are presented in Ref.~\cite{Kniehl:2012mn} for the unpolarized top quark decays. The NLO differential decay rates for the polarized top quarks are analytically calculated in~\cite{Nejad:2013fba,MoosaviNejad:2011yp}.
Although, in the above relation the factorization ($\mu_F$) and renormalization ($\mu_R$) scales are arbitrary but we here set them to $\mu_R = \mu_F = m_t = 172.9$ GeV.
Adopting $m_b=4.75$ GeV and $m_W=80.39$ GeV, in Fig.~\ref{fig1} we studied the energy distribution of $D^{*\pm}$-mesons in the polarized top decay at NLO (solid line). Using the NLO FFs of $(b,g)\to D^{*+}$ extracted in Ref.~\cite{Kneesch:2007ey}, we also studied the same quantity (dashed line) in Fig.~\ref{fig1}. As is seen, in comparison with the KKKS08 result, our FFs show a reduction for the energy spectrum at the peak position. In Fig.~\ref{fig2}, we have presented our NLO predictions for the energy distribution of $D^{*\pm}$-mesons produced through the polarized (dashed line) and unpolarized (dots) top decays.


To have a detailed insight for the size of NNLO corrections, in this plot we have also studied the same distribution at NNLO for the polarized top decays (solid line) and unpolarized ones (dot-dashed). However, these NNLO predictions are not completely correct because for a reliable prediction at this order of perturbative QCD one also needs to have the NNLO parton-level differential decay rates $d\Gamma/dx_i$ to convolute with the NNLO FFs, see Eq.~\eqref{eq:master}. These quantities have not been yet calculated.
The study of the $x_D$ distribution (i.e. $d\Gamma/dx_D$) of the dominant decay mode $t\to D^{*\pm}W^++X$ at the LHC, as a formidable top factory, will enable us to deepen our understanding of the nonperturbative aspects of $D^{*\pm}$-meson formation by hadronization and to pin down the $b\to D^{*\pm}$ and $g\to D^{*\pm}$ FFs. By measuring the $x_D$ distribution of polarized top decays, the $b/g\to D^{*\pm}$ FFs can be constrained event further.


%
\section{Summary and Conclusions} \label{sec:conclusion}
%

Let us now come to our summary and conclusions. We have determined the nonperturbative FFs of partons into the $D^{*\pm}$-meson at NLO perturbative QCD and, for the first time, at NNLO one from global analyses of single-inclusive electron-positron annihilation. Our analyses are based on the ZM-VFN scheme in which all quarks are treated as massless partons.
Our phenomenological analyses (called as {\tt SKM18} analyses) to achieve the FFs of $D^{*\pm}$-meson is significant in, at least, two major respects. Firstly, we applied all SIA experimental data as much as possible including most of the data from ALEPH and OPAL Collaborations. Secondly, we considered the NNLO accuracy in our global fit using the public {\tt APFEL} code. According to the $\chi^{2}$ values extracted from our optimum fits, increasing the accuracies of theoretical QCD analysis up to NNLO slightly decreases the value of $\chi^2$. In addition, we found that the experimental uncertainties for the $D^{*\pm}$-FFs and SIA cross sections are similar in size both for the NLO and NNLO approximations. We  have also studied the variation effects of the renormalization and factorization scales considering  $Q/2\le \mu \le 2Q$ where we set $ \mu_r=\mu_f=\mu$. The obtained results show that our calculations at NNLO come with much smaller theoretical uncertainties relative to the NLO calculations which reflects the stability of our NNLO analysis. These findings are significantly in agreements with previous results reported in the literature.

As is well-known in global QCD analysis of FFs at NLO, some phenomenological collaborations have included single-inclusive $D^{*\pm}$-meson production in electron-positron annihilation, hadron-hadron collisions, and in-jet fragmentation in proton-proton scattering. Unfortunately, the NNLO expressions of Wilson coefficients for the processes of hadron-hadron collision and proton-proton scattering are not yet available. Therefore, we restricted our analyses to the SIA experimental data.  
Despite to a few number of experimental data available for $D^{*\pm}$-meson, the {\tt SKM18} theoretical predictions are in reasonable agreements with the experimental data.

We also applied the {\tt SKM18} FFs to make our predictions for the energy spectrum of $D^{*\pm}$-mesons produced in polarized and unpolarized top quark decays. At the LHC, this study can be considered as a channel to indirect search for the top properties and, specifically, it enables us to deepen our understanding of the nonperturbative aspects of $D^{*\pm}$-meson formation and to pin down the $(b, g)\to D^{*\pm}$ FFs. 

As was mentioned, including the mass effects of heavy quarks and meson in the theoretical QCD analysis leads to the significant information about our understanding of FFs, specially at the small-$z$ region. The analyses including the meson and quark masses are reserved for future work. More detail on these effects will be given in our upcoming paper. Although the current study is based on a few sample of datasets, the findings are in good agreements with all experimental data as well as the results in literature. We hope that our research will serve as a base for future studies on $D^{*\pm}$-meson FFs. However, we should emphasize that a number of future studies using the experimental data from colliders are strongly recommended.

\clearpage

%
\begin{acknowledgments}
%

We gratefully acknowledge the help and technical assistance provided by Valerio Bertone  and Muhammad Goharipour. We also would like to thank Daniele P. Anderle for providing us with the best fit of {\tt AKSRV17} NNLO FFs.
Authors thank School of Particles and Accelerators, Institute for Research in Fundamental Sciences (IPM) for financial support of this project.
Hamzeh Khanpour also is thankful the University of Science and Technology of Mazandaran for financial support provided for this research.
Mohammad Moosavi Nejad is grateful Yaz university for financial support for this project.

\end{acknowledgments}
%



\begin{thebibliography}{}



\bibitem{Bertone:2017tyb}
V.~Bertone {\it et al.} [NNPDF Collaboration],
``A determination of the fragmentation functions of pions, kaons, and protons with faithful uncertainties,''
Eur.\ Phys.\ J.\ C {\bf 77} (2017) no.8,  516
[arXiv:1706.07049 [hep-ph]].





\bibitem{Anderle:2015lqa} 
D.~P.~Anderle, F.~Ringer and M.~Stratmann,
``Fragmentation Functions at Next-to-Next-to-Leading Order Accuracy,''
Phys.\ Rev.\ D {\bf 92}, no. 11, 114017 (2015)
[arXiv:1510.05845 [hep-ph]].




\bibitem{Ethier:2017zbq} 
J.~J.~Ethier, N.~Sato and W.~Melnitchouk,
``First simultaneous extraction of spin-dependent parton distributions and fragmentation functions from a global QCD analysis,''
Phys.\ Rev.\ Lett.\  {\bf 119}, no. 13, 132001 (2017)
[arXiv:1705.05889 [hep-ph]].

\bibitem{MoosaviNejad:2016qdx}
S.~M.~Moosavi Nejad and P.~Sartipi Yarahmadi,
Eur.\ Phys.\ J.\ A {\bf 52} (2016) no.10,  315
doi:10.1140/epja/i2016-16315-7
[arXiv:1609.07422 [hep-ph]].





\bibitem{Nocera:2017zge} 
E.~R.~Nocera and M.~Ubiali,
``Constraining the gluon PDF at large x with LHC data,''
arXiv:1709.09690 [hep-ph].




\bibitem{Hou:2017khm} 
T.~J.~Hou {\it et al.},
``CT14 Intrinsic Charm Parton Distribution Functions from CTEQ-TEA Global Analysis,''
arXiv:1707.00657 [hep-ph].









\bibitem{Eskola:2016oht} 
K.~J.~Eskola, P.~Paakkinen, H.~Paukkunen and C.~A.~Salgado,
``EPPS16: Nuclear parton distributions with LHC data,''
Eur.\ Phys.\ J.\ C {\bf 77}, no. 3, 163 (2017)
[arXiv:1612.05741 [hep-ph]].





\bibitem{Frankfurt:2016qca} 
L.~Frankfurt, V.~Guzey and M.~Strikman,
``Dynamical model of antishadowing of the nuclear gluon distribution,''
Phys.\ Rev.\ C {\bf 95}, no. 5, 055208 (2017)
[arXiv:1612.08273 [hep-ph]].




\bibitem{Goharipour:2017uic} 
M.~Goharipour and H.~Mehraban,
``Study of isolated prompt photon production in $ p $-Pb collisions for the ALICE kinematics,''
Phys.\ Rev.\ D {\bf 95}, no. 5, 054002 (2017)
[arXiv:1702.05738 [hep-ph]].













\bibitem{Slovak:2017xdc} 
R.~Slovak [ATLAS Collaboration],
``Jet Fragmentation in pp , p + Pb and Pb + Pb collisions in the ATLAS detector,''
Nucl.\ Phys.\ A {\bf 967}, 504 (2017).





\bibitem{Anderle:2017cgl} 
D.~P.~Anderle, T.~Kaufmann, M.~Stratmann, F.~Ringer and I.~Vitev,
``Using hadron-in-jet data in a global analysis of $D^{*}$ fragmentation functions,''
Phys.\ Rev.\ D {\bf 96}, no. 3, 034028 (2017)
[arXiv:1706.09857 [hep-ph]].





\bibitem{DGLAP}
V.~N.~Gribov and L.~N.~Lipatov,
Sov.\ J.\ Nucl.\ Phys.\  {\bf 15} (1972) 438
[Yad.\ Fiz.\  {\bf 15} (1972) 781]; 
G.~Altarelli and G.~Parisi,
Nucl.\ Phys.\ B {\bf 126} (1977) 298; 
Sov.\ Phys.\ JETP {\bf 46} (1977) 641
[Zh.\ Eksp.\ Teor.\ Fiz.\  {\bf 73} (1977) 1216].
	




\bibitem{Zarei:2015jvh} 
M.~Zarei, F.~Taghavi-Shahri, S.~Atashbar Tehrani and M.~Sarbishei,
``Fragmentation functions of the pion, kaon, and proton in the NLO approximation: Laplace transform approach,''
Phys.\ Rev.\ D {\bf 92}, no. 7, 074046 (2015)
[arXiv:1601.02815 [hep-ph]].





\bibitem{Boroun:2015aya} 
G.~R.~Boroun, T.~Osati and S.~Zarrin,
``An Approximation Approach to the Evolution of the Fragmentation Function,''
Int.\ J.\ Theor.\ Phys.\  {\bf 54}, no. 10, 3831 (2015).




\bibitem{Boroun:2016zql} 
G.~R.~Boroun, S.~Zarrin and S.~Dadfar,
``Laplace method for the evolution of the fragmentation function of B$_c$ mesons,''
Nucl.\ Phys.\ A {\bf 953}, 21 (2016).



\bibitem{Soleymaninia:2013cxa} 
M.~Soleymaninia, A.~N.~Khorramian, S.~M.~Moosavi Nejad and F.~Arbabifar,
``Determination of pion and kaon fragmentation functions including spin asymmetries data in a global analysis,''
Phys.\ Rev.\ D {\bf 88}, no. 5, 054019 (2013)
Addendum: [Phys.\ Rev.\ D {\bf 89}, no. 3, 039901 (2014)]
[arXiv:1306.1612 [hep-ph]].
	
	
	


\bibitem{Nejad:2015fdh} 
S.~M.~Moosavi Nejad, M.~Soleymaninia and A.~Maktoubian,
``Proton fragmentation functions considering finite-mass corrections,''
Eur.\ Phys.\ J.\ A {\bf 52}, no. 10, 316 (2016)
[arXiv:1512.01855 [hep-ph]].


	


\bibitem{Albino:2008fy} 
S.~Albino, B.~A.~Kniehl and G.~Kramer,
``AKK Update: Improvements from New Theoretical Input and Experimental Data,''
Nucl.\ Phys.\ B {\bf 803}, 42 (2008)
[arXiv:0803.2768 [hep-ph]].





\bibitem{Leader:2015hna} 
E.~Leader, A.~V.~Sidorov and D.~B.~Stamenov,
``Determination of the fragmentation functions from an NLO QCD analysis of the HERMES data on pion multiplicities,''
Phys.\ Rev.\ D {\bf 93}, no. 7, 074026 (2016)
[arXiv:1506.06381 [hep-ph]].





\bibitem{deFlorian:2014xna} 
D.~de Florian, R.~Sassot, M.~Epele, R.~J.~Hernández-Pinto and M.~Stratmann,
``Parton-to-Pion Fragmentation Reloaded,''
Phys.\ Rev.\ D {\bf 91}, no. 1, 014035 (2015)
[arXiv:1410.6027 [hep-ph]].




\bibitem{deFlorian:2007aj} 
D.~de Florian, R.~Sassot and M.~Stratmann,
``Global analysis of fragmentation functions for pions and kaons and their uncertainties,''
Phys.\ Rev.\ D {\bf 75}, 114010 (2007)
[hep-ph/0703242 [HEP-PH]].




\bibitem{deFlorian:2007hc} 
D.~de Florian, R.~Sassot and M.~Stratmann,
Phys.\ Rev.\ D {\bf 76}, 074033 (2007)
[arXiv:0707.1506 [hep-ph]].





\bibitem{Hirai:2007cx} 
M.~Hirai, S.~Kumano, T.-H.~Nagai and K.~Sudoh,
``Determination of fragmentation functions and their uncertainties,''
Phys.\ Rev.\ D {\bf 75}, 094009 (2007)
[hep-ph/0702250].




\bibitem{Sato:2016wqj} 
N.~Sato, J.~J.~Ethier, W.~Melnitchouk, M.~Hirai, S.~Kumano and A.~Accardi,
``First Monte Carlo analysis of fragmentation functions from single-inclusive $e^+ e^-$ annihilation,''
Phys.\ Rev.\ D {\bf 94}, no. 11, 114004 (2016)
[arXiv:1609.00899 [hep-ph]].




\bibitem{Pumplin:2000vx} 
J.~Pumplin, D.~R.~Stump and W.~K.~Tung,
``Multivariate fitting and the error matrix in global analysis of data,''
Phys.\ Rev.\ D {\bf 65}, 014011 (2001)
[hep-ph/0008191].





\bibitem{Sato:2016tuz} 
N.~Sato {\it et al.} [Jefferson Lab Angular Momentum Collaboration],
``Iterative Monte Carlo analysis of spin-dependent parton distributions,''
Phys.\ Rev.\ D {\bf 93}, no. 7, 074005 (2016)
[arXiv:1601.07782 [hep-ph]].


\bibitem{Ball:2017nwa} 
R.~D.~Ball {\it et al.} [NNPDF Collaboration],
``Parton distributions from high-precision collider data,''
Eur.\ Phys.\ J.\ C {\bf 77}, no. 10, 663 (2017)
[arXiv:1706.00428 [hep-ph]].




\bibitem{Gauld:2015yia} 
R.~Gauld, J.~Rojo, L.~Rottoli and J.~Talbert,
``Charm production in the forward region: constraints on the small-x gluon and backgrounds for neutrino astronomy,''
JHEP {\bf 1511}, 009 (2015)
[arXiv:1506.08025 [hep-ph]].





\bibitem{Garzelli:2015psa}
M.~V.~Garzelli, S.~Moch and G.~Sigl,
JHEP {\bf 1510} (2015) 115;
A.~Bhattacharya, R.~Enberg, M.~H.~Reno, I.~Sarcevic and A.~Stasto,
JHEP {\bf 1506} (2015) 110.


\bibitem{Gauld:2015kvh} 
  R.~Gauld, J.~Rojo, L.~Rottoli, S.~Sarkar and J.~Talbert,
  ``The prompt atmospheric neutrino flux in the light of LHCb,''
  JHEP {\bf 1602}, 130 (2016)
  [arXiv:1511.06346 [hep-ph]].



\bibitem{Kang:2016ofv} 
Z.~B.~Kang, F.~Ringer and I.~Vitev,
``Effective field theory approach to open heavy flavor production in heavy-ion collisions,''
JHEP {\bf 1703}, 146 (2017)
[arXiv:1610.02043 [hep-ph]].




\bibitem{Kneesch:2007ey} 
T.~Kneesch, B.~A.~Kniehl, G.~Kramer and I.~Schienbein,
``Charmed-meson fragmentation functions with finite-mass corrections,''
Nucl.\ Phys.\ B {\bf 799}, 34 (2008)
[arXiv:0712.0481 [hep-ph]].




\bibitem{Binnewies:1998vm} 
J.~Binnewies, B.~A.~Kniehl and G.~Kramer,
``Inclusive $B$ meson production in $e^{+} e^{-}$ and $p \bar{p}$ collisions,''
Phys.\ Rev.\ D {\bf 58}, 034016 (1998)
[hep-ph/9802231].




\bibitem{Barate:1999bg} 
R.~Barate {\it et al.} [ALEPH Collaboration],
``Study of charm production in Z decays,''
Eur.\ Phys.\ J.\ C {\bf 16}, 597 (2000)
[hep-ex/9909032].
  
  
  
  
  
\bibitem{Ackerstaff:1997ki} 
K.~Ackerstaff {\it et al.} [OPAL Collaboration],
``Measurement of f(c ---> D*+ X), f(b ---> D*+ X) and Gamma (c anti-c) / Gamma (hadronic) using D*+- mesons,''
Eur.\ Phys.\ J.\ C {\bf 1}, 439 (1998)
[hep-ex/9708021].






\bibitem{Martin:2002aw} 
A.~D.~Martin, R.~G.~Roberts, W.~J.~Stirling and R.~S.~Thorne,
``Uncertainties of predictions from parton distributions. 1: Experimental errors,''
Eur.\ Phys.\ J.\ C {\bf 28}, 455 (2003)
[hep-ph/0211080].










\bibitem{Pumplin:2001ct} 
J.~Pumplin, D.~Stump, R.~Brock, D.~Casey, J.~Huston, J.~Kalk, H.~L.~Lai and W.~K.~Tung,
``Uncertainties of predictions from parton distribution functions. 2. The Hessian method,''
Phys.\ Rev.\ D {\bf 65}, 014013 (2001)
[hep-ph/0101032].





\bibitem{Martin:2009iq} 
A.~D.~Martin, W.~J.~Stirling, R.~S.~Thorne and G.~Watt,
``Parton distributions for the LHC,''
Eur.\ Phys.\ J.\ C {\bf 63}, 189 (2009)
[arXiv:0901.0002 [hep-ph]].




\bibitem{Collins:1998rz} 
J.~C.~Collins,
``Hard scattering factorization with heavy quarks: A General treatment,''
Phys.\ Rev.\ D {\bf 58}, 094002 (1998)
[hep-ph/9806259].





\bibitem{Gorishnii:1990vf} 
S.~G.~Gorishnii, A.~L.~Kataev and S.~A.~Larin,
``The $O(\alpha^{3}_{s})$-corrections to $\sigma_{tot}(e^{+}e^{-}\rightarrow hadrons)$ and $\Gamma(\tau^{-} \rightarrow \nu_{\tau} + hadrons)$ in QCD,''
Phys.\ Lett.\ B {\bf 259}, 144 (1991).




\bibitem{Rijken:1996vr} 
P.~J.~Rijken and W.~L.~van Neerven,
``O (alpha-s**2) contributions to the longitudinal fragmentation function in e+ e- annihilation,''
Phys.\ Lett.\ B {\bf 386}, 422 (1996)
[hep-ph/9604436].
 
 
 
\bibitem{Rijken:1996ns} 
P.~J.~Rijken and W.~L.~van Neerven,
``Higher order QCD corrections to the transverse and longitudinal fragmentation functions in electron - positron annihilation,''
Nucl.\ Phys.\ B {\bf 487}, 233 (1997)
[hep-ph/9609377].
 
 
 
 
\bibitem{Mitov:2006wy} 
A.~Mitov and S.~O.~Moch,
``QCD Corrections to Semi-Inclusive Hadron Production in Electron-Positron Annihilation at Two Loops,''
Nucl.\ Phys.\ B {\bf 751}, 18 (2006)
[hep-ph/0604160].





\bibitem{Kniehl:2006mw} 
B.~A.~Kniehl and G.~Kramer,
``Charmed-hadron fragmentation functions from CERN LEP1 revisited,''
Phys.\ Rev.\ D {\bf 74}, 037502 (2006)
[hep-ph/0607306].
  
  
  
  
\bibitem{Binnewies:1997xq} 
J.~Binnewies, B.~A.~Kniehl and G.~Kramer,
``Predictions for D*+- - photoproduction at HERA with new fragmentation functions from LEP-1,''
Phys.\ Rev.\ D {\bf 58}, 014014 (1998)
[hep-ph/9712482].
  
    
  
  
  
  
\bibitem{Artuso:2004pj} 
M.~Artuso {\it et al.} [CLEO Collaboration],
``Charm meson spectra in $e^{+} e^{-}$ annihilation at 10.5-GeV c.m.e.,''
Phys.\ Rev.\ D {\bf 70}, 112001 (2004)
[hep-ex/0402040].





  
 
\bibitem{Seuster:2005tr} 
R.~Seuster {\it et al.} [Belle Collaboration],
``Charm hadrons from fragmentation and B decays in e+ e- annihilation at s**(1/2) = 10.6-GeV,''
Phys.\ Rev.\ D {\bf 73}, 032002 (2006)
[hep-ex/0506068].





\bibitem{Belle} 
http://durpdg.dur.ac.uk/view/ins686014.





\bibitem{Cacciari:2005uk} 
M.~Cacciari, P.~Nason and C.~Oleari,
``A Study of heavy flavored meson fragmentation functions in e+ e- annihilation,''
JHEP {\bf 0604}, 006 (2006)
[hep-ph/0510032].


  
  

\bibitem{Patrignani:2016xqp} 
C.~Patrignani {\it et al.} [Particle Data Group],
``Review of Particle Physics,''
Chin.\ Phys.\ C {\bf 40}, no. 10, 100001 (2016).
  
 
  




\bibitem{James:1994vla} 
F.~James,
``MINUIT Function Minimization and Error Analysis:  Reference Manual Version 94.1,''
CERN-D-506, CERN-D506.







\bibitem{Metz:2016swz} 
A.~Metz and A.~Vossen,
``Parton Fragmentation Functions,''
Prog.\ Part.\ Nucl.\ Phys.\  {\bf 91}, 136 (2016)
[arXiv:1607.02521 [hep-ex]].





\bibitem{Anderle:2016czy} 
D.~P.~Anderle, T.~Kaufmann, M.~Stratmann and F.~Ringer,
``Fragmentation Functions Beyond Fixed Order Accuracy,''
Phys.\ Rev.\ D {\bf 95}, no. 5, 054003 (2017)
[arXiv:1611.03371 [hep-ph]].





\bibitem{deFlorian:2017lwf} 
D.~de Florian, M.~Epele, R.~J.~Hernandez-Pinto, R.~Sassot and M.~Stratmann,
``Parton-to-Kaon Fragmentation Revisited,''
Phys.\ Rev.\ D {\bf 95}, no. 9, 094019 (2017)
[arXiv:1702.06353 [hep-ph]].



\bibitem{Gao:2017yyd} 
J.~Gao, L.~Harland-Lang and J.~Rojo,
``The Structure of the Proton in the LHC Precision Era,''
arXiv:1709.04922 [hep-ph].








\bibitem{Shoeibi:2017lrl} 
S.~Shoeibi, H.~Khanpour, F.~Taghavi-Shahri and K.~Javidan,
``Determination of neutron fracture functions from a global QCD analysis of the leading neutron production at HERA,''
Phys.\ Rev.\ D {\bf 95}, no. 7, 074011 (2017)
[arXiv:1703.04369 [hep-ph]].





\bibitem{Ball:2016neh} 
R.~D.~Ball {\it et al.} [NNPDF Collaboration],
``A Determination of the Charm Content of the Proton,''
Eur.\ Phys.\ J.\ C {\bf 76}, no. 11, 647 (2016)
[arXiv:1605.06515 [hep-ph]].




\bibitem{Ubiali:2014bva} 
M.~Ubiali [NNPDF Collaboration],
``Towards the NNPDF3.0 parton set for the second LHC run,''
PoS DIS {\bf 2014}, 041 (2014)
[arXiv:1407.3122 [hep-ph]].





\bibitem{Ball:2013tyh} 
R.~D.~Ball {\it et al.} [NNPDF Collaboration],
``Polarized Parton Distributions at an Electron-Ion Collider,''
Phys.\ Lett.\ B {\bf 728}, 524 (2014)
[arXiv:1310.0461 [hep-ph]].






\bibitem{Nocera:2014gqa} 
E.~R.~Nocera {\it et al.} [NNPDF Collaboration],
``A first unbiased global determination of polarized PDFs and their uncertainties,''
Nucl.\ Phys.\ B {\bf 887}, 276 (2014)
[arXiv:1406.5539 [hep-ph]].




\bibitem{Shoeibi:2017zha} 
S.~Shoeibi, F.~Taghavi-Shahri, H.~Khanpour and K.~Javidan,
``Phenomenology of leading nucleon production in $ep$ collisions at HERA in the framework of fracture functions,''
arXiv:1710.06329 [hep-ph].



\bibitem{Shahri:2016uzl} 
F.~Taghavi-Shahri, H.~Khanpour, S.~Atashbar Tehrani and Z.~Alizadeh Yazdi,
``Next-to-next-to-leading order QCD analysis of spin-dependent parton distribution functions and their uncertainties: Jacobi polynomials approach,''
Phys.\ Rev.\ D {\bf 93}, no. 11, 114024 (2016)
[arXiv:1603.03157 [hep-ph]].




\bibitem{Khanpour:2017fey} 
H.~Khanpour, S.~T.~Monfared and S.~Atashbar Tehrani,
``Study of spin-dependent structure functions of $^3{\rm He}$ and $^3{\rm H}$ at NNLO approximation and corresponding nuclear corrections,''
Phys.\ Rev.\ D {\bf 96}, no. 7, 074037 (2017)
[arXiv:1710.05747 [hep-ph]].




\bibitem{Khanpour:2017cha} 
H.~Khanpour, S.~T.~Monfared and S.~Atashbar Tehrani,
``Nucleon spin structure functions at NNLO in the presence of target mass corrections and higher twist effects,''
Phys.\ Rev.\ D {\bf 95}, no. 7, 074006 (2017)
[arXiv:1703.09209 [hep-ph]].





\bibitem{Arbabifar:2013tma} 
F.~Arbabifar, A.~N.~Khorramian and M.~Soleymaninia,
``QCD analysis of polarized DIS and the SIDIS asymmetry world data and light sea-quark decomposition,''
Phys.\ Rev.\ D {\bf 89}, no. 3, 034006 (2014)
[arXiv:1311.1830 [hep-ph]].




\bibitem{Khanpour:2016pph} 
H.~Khanpour and S.~Atashbar Tehrani,
``Global Analysis of Nuclear Parton Distribution Functions and Their Uncertainties at Next-to-Next-to-Leading Order,''
Phys.\ Rev.\ D {\bf 93}, no. 1, 014026 (2016)
[arXiv:1601.00939 [hep-ph]].




\bibitem{MoosaviNejad:2016ebo} 
S.~M.~Moosavi Nejad, H.~Khanpour, S.~Atashbar Tehrani and M.~Mahdavi,
``QCD analysis of nucleon structure functions in deep-inelastic neutrino-nucleon scattering: Laplace transform and Jacobi polynomials approach,''
Phys.\ Rev.\ C {\bf 94}, no. 4, 045201 (2016)
[arXiv:1609.05310 [hep-ph]].





\bibitem{Khanpour:2016uxh} 
H.~Khanpour, A.~Mirjalili and S.~Atashbar Tehrani,
``Analytic derivation of the next-to-leading order proton structure function $F_2^p(x, Q^2)$ based on the Laplace transformation,''
Phys.\ Rev.\ C {\bf 95}, no. 3, 035201 (2017)
[arXiv:1601.03508 [hep-ph]].





\bibitem{Bertone:2013vaa} 
V.~Bertone, S.~Carrazza and J.~Rojo,
``APFEL: A PDF Evolution Library with QED corrections,''
Comput.\ Phys.\ Commun.\  {\bf 185}, 1647 (2014)
[arXiv:1310.1394 [hep-ph]].




 
\bibitem{Nocera:2017gbk} 
E.~R.~Nocera,
``Fragmentation functions of charged hadrons,''
arXiv:1709.03400 [hep-ph].
 
 
   
  
\bibitem{KKKS08code} 
The KKKS08 code we use was downloaded from http://lapth.cnrs.fr/ffgenerator/.




\bibitem{Kniehl:2012mn} 
B.~A.~Kniehl, G.~Kramer and S.~M.~Moosavi Nejad,
``Bottom-Flavored Hadrons from Top-Quark Decay at Next-to-Leading order in the General-Mass Variable-Flavor-Number Scheme,''
Nucl.\ Phys.\ B {\bf 862}, 720 (2012)
[arXiv:1205.2528 [hep-ph]].





\bibitem{Nejad:2013fba} 
S.~M.~Moosavi Nejad,
``Energy spectrum of bottom- and charmed-flavored mesons from polarized top quark decay $t(\uparrow)\to W^++B/D+X$ at $O(\alpha_s)$,''
Phys.\ Rev.\ D {\bf 88}, no. 9, 094011 (2013)
[arXiv:1310.5686 [hep-ph]].


\bibitem{MoosaviNejad:2011yp} 
S.~M.~Moosavi Nejad,
Phys.\ Rev.\ D {\bf 85}, 054010 (2012).









\end{thebibliography}
\end{document}